\documentclass[aps,prb,showpacs,twocolumn,superscriptaddress]{revtex4}

\def\be{\begin{equation}}
\def\ee{\end{equation}}

\def\vec{\mathbf}
\def\mc{\mathcal}

\usepackage{amssymb}
\usepackage[dvips]{graphicx}
\usepackage{epsfig}
\usepackage{hyperref}

\begin{document}
\title{Highly Frustrated Magnetic Clusters: The kagom\'e on a sphere}
\author{Ioannis Rousochatzakis}
\email{ioannis.rousochatzakis@epfl.ch}
\affiliation{Institut de th\'eorie des ph\'enom\`enes physiques,
Ecole polytechnique f\'ed\'erale de Lausanne,\\ CH-1015 Lausanne, Switzerland}
\author{Andreas M. L\"auchli}
%\email{andreas.laeuchli@epfl.ch}
\affiliation{Institut Romand de Recherche Num\'erique en Physique de Mat\'eriaux (IRRMA),\\
CH-1015 Lausanne, Switzerland}
\author{Fr\'ed\'eric Mila}
%\email{frederic.mila@epfl.ch}
\affiliation{Institut de th\'eorie des ph\'enom\`enes physiques,
Ecole polytechnique f\'ed\'erale de Lausanne,\\ CH-1015 Lausanne, Switzerland}

\begin{abstract}
We present a detailed study of the low-energy excitations of two existing finite-size
realizations of the planar kagom\'e Heisenberg antiferromagnet on the sphere,
the cuboctahedron and the icosidodecahedron. 
%An adaptation of the standard exact diagonalization method which exploits the full point group symmetries has been employed. 
After highlighting a number of
special spectral features (such as the presence of low-lying singlets below the first triplet
and the existence of localized magnons) we focus on two major issues. The first concerns the nature of the excitations
above the plateau phase at $1/3$ of the saturation magnetization $M_s$.
Our exact diagonalizations for the $s=1/2$ icosidodecahedron reveal that the low-lying plateau states
are adiabatically connected to the degenerate collinear ``up-up-down'' ground states of the Ising point,
at the same time being well isolated from higher excitations. A complementary physical picture emerges 
from the derivation of an effective quantum dimer model which reveals the central role of the topology and the 
intrinsic spin $s$. We also give a prediction for the low energy excitations and thermodynamic properties of 
the spin $s=5/2$ icosidodecahedron Mo$_{72}$Fe$_{30}$. 
In the second part we focus on the low-energy spectra of the $s>1/2$ Heisenberg model in view of interpreting 
the broad inelastic neutron scattering response reported for Mo$_{72}$Fe$_{30}$. 
To this end we demonstrate the simultaneous presence of several broadened low-energy ``towers of states'' 
or ``rotational bands'' which arise from the large discrete spatial degeneracy of the classical ground states, 
a generic feature of highly frustrated clusters. 
This semiclassical interpretation is further corroborated by their striking symmetry pattern 
which is shown, by an independent group theoretical analysis, to be a characteristic 
fingerprint of the classical coplanar ground states.
\end{abstract}
\date{\today}

\maketitle

\section{Introduction}\label{Intro.Sec}
The field of highly frustrated magnetism has received a growing theoretical and experimental
interest in recent years~\cite{Misquish_Review_2D_AFM,Richter_chapter}.
One of the central motifs in the planar kagom\'e and similarly frustrated Heisenberg
antiferromagnets (AFM's) which readily differentiates them from unfrustrated (e.g. collinear) ones,
is the proliferation of an extensive family of low-energy singlets below the lowest triplet
excitation~\cite{Lecheminant_kagome,Waldtmann}.
One interpretation for the origin of these singlets has emerged from Resonating Valence Bond (RVB) type of
arguments~\cite{Mila} for the $s=1/2$ kagom\'e AFM.
For higher spins, purely classical considerations assert that the singlets stem from the splitting by
quantum fluctuations of the extensively degenerate family of N\'eel ordered (3-sublattice) ground
states~\cite{Moessner_Review_Class_Deg}.
Both interpretations rest on the notion of a local degeneracy which stems from the frustrated corner-sharing
topology of these lattices. In this regard, it appears that the proliferation of singlets is only one
particular manifestation of this local degeneracy since similarly dense low-energy excitations are
manifested in the whole magnetization range.

On the other hand, some understanding for the ground state itself has been established.
Exact Diagonalization (ED) results suggest that the ground state of the $s=1/2$ kagom\'e AFM
is a disordered spin liquid with a very small spin gap~\cite{Lecheminant_kagome,Waldtmann} (if any).
For $s>1/2$, semi-classical approaches predict that an extensive subset of coplanar
states is first selected in $1/s$ while the $\sqrt{3}\times\sqrt{3}$ ordered state is stabilized
in higher orders through the order-by-disorder mechanism~\cite{Chubukov,ChanHenley}.
In a magnetic field, the ground state may exhibit a number of interesting phases.
These include the presence of an extensively degenerate family of localized magnons which result in
macroscopic magnetization jumps at the saturation field, as well as the stabilization of spin gaps
and the associated fractional magnetization plateaux.
For a first understanding of the nature of these plateaux a perturbative expansion around
the degenerate Ising point was first employed by Cabra \textit{et al.}~\cite{Cabra} for the kagom\'e.
This approach was recently extended by Bergman \textit{et al.}~\cite{Bergman1,Bergman2}
to other frustrated systems, such as the pyrochlore AFM.
Here, the anisotropy terms are treated perturbatively, and the emerging splitting of the degenerate
Ising manifold is effectively cast into a Quantum Dimer Model (QDM) on the dual lattice.

At the same time, it is well known that some precursors of the excitation spectra of frustrated and
unfrustrated AFM's are already embodied in the spectra of small system sizes
(see for instance Ref.~\onlinecite{Lhuillier_Towers}).
It has come therefore with no surprise that a number of phenomena that are manifest in kagom\'e-like AFM's
have also emerged in the research field of highly frustrated nanomagnets~\cite{GSV,Schnack_Review,
Christian_Meta,Fe30_Plateau,Schmidt_Singlets,Schnack_ind_magn_MMs,Schnack_caloric}.
These are realizations of zero-dimensional molecular-size magnets which consist of a finite number of
strongly interacting transition metal ions, with the isotropic Heisenberg exchange being the dominant energy
term. Thus, in addition to their great relevance in the context of nanomagnetism and the growing interest
for potential applications in quantum computing~\cite{Appl_QC}, information storage~\cite{Appl_MS} and magnetic
imaging~\cite{Appl_MI}, molecular nanomagnets can also provide a suitable platform for addressing theoretical
questions and testing ideas from the more general context of frustrated magnetism.

\begin{figure}[!b]
\centering
\includegraphics*[width=0.4\linewidth]{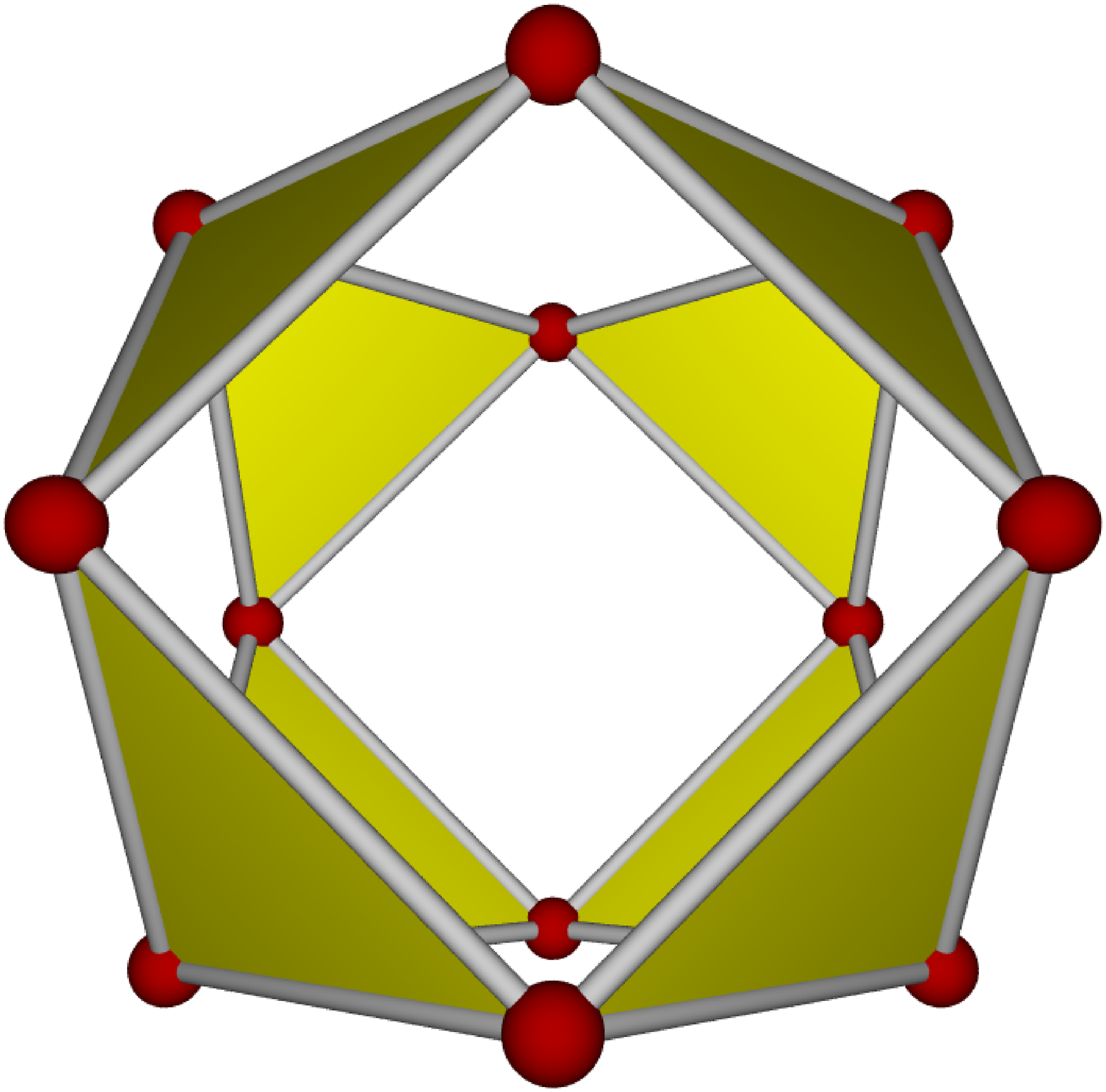}
\hspace{0.1\linewidth}\includegraphics*[width=0.4\linewidth]{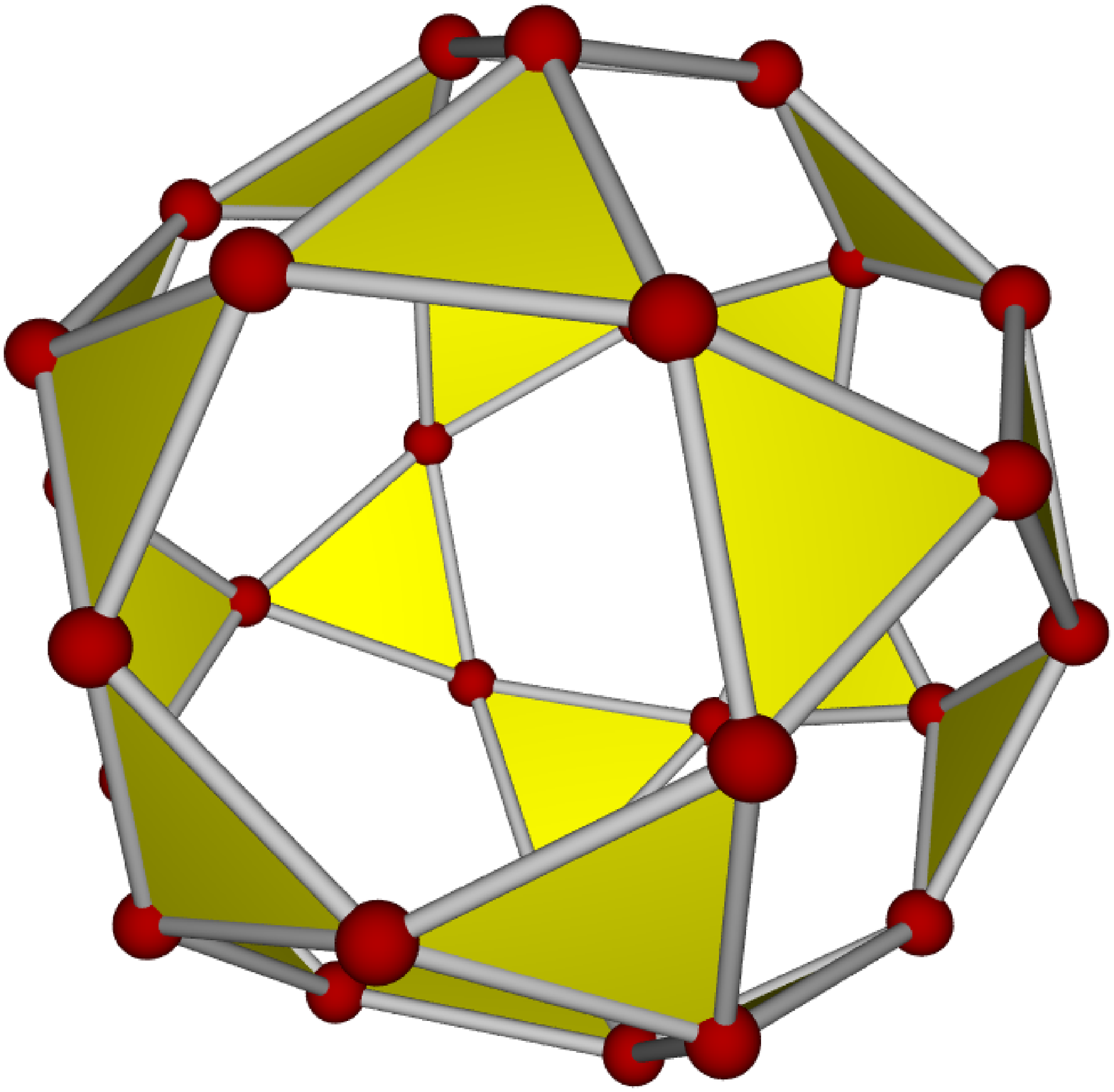}
\caption{(Color online) Schematic representation of the cuboctahedron (left) and the icosidodecahedron (right).
The first consists of 12 vertices, 24 edges, 6 square and 8 triangular faces,
while the latter consists of 30 vertices, 60 edges, 12 pentagons and 20 corner-sharing triangles.}
\label{CubocIcosi.Fig}
\end{figure}

In this work, we focus on two magnetic molecule realizations of the Heisenberg kagom\'e AFM on the sphere.
The first consists of $8$ corner-sharing triangles and is realized in the Cu$_{12}$La$_8$~\cite{Cu12} cluster 
with $12$ Cu$^{2+}$ $s=1/2$ ions occupying the vertices of a symmetric cuboctahedron (cf.Fig.~\ref{CubocIcosi.Fig}). 
The spin topology of this cluster is identical to the $12$-site kagom\'e wrapped on a torus 
(cf. Fig.~\ref{CubocOnThePlane.Fig}). The second cluster is one of the largest frustrated molecules synthesized to date, 
namely the giant Keplerate Mo$_{72}$Fe$_{30}$ system~\cite{Muller_Fe30}. This features an array of thirty $s=5/2$ Fe$^{3+}$ 
ions occupying the vertices of twenty corner-sharing triangles spanning an almost perfect icosidodecahedron 
(cf.Fig.~\ref{CubocIcosi.Fig}). Interestingly, its quantum $s=1/2$ analogue, Mo$_{72}$V$_{30}$, consisting of V$^{4+}$ ions
has also been synthesized quite recently~\cite{V30_1,V30_2}. 
We may note here that the cuboctahedron and the icosidodecahedron can be thought of as two existing positive curvature 
(with $n=$ 4 and 5 respectively) counterparts of Elser and Zeng's~\cite{ElserZeng} generalization of the kagom\'e 
structure on the hyperbolic plane where each hexagon is replaced by a polygon of $n$ sides with $n>6$. 

Among the above highly frustrated clusters, Mo$_{72}$Fe$_{30}$ has been the most investigated so far, both theoretically 
and experimentally. The exchange interactions in Mo$_{72}$Fe$_{30}$ are quite small, $J/k_B\simeq 1.57$ K~\cite{Muller_Fe30},
and this has allowed for the experimental observation of a $M=M_s/3$ plateau at $H\simeq 5.9$ Tesla 
which has been explained classically by Schr\"oder \textit{et al.}~\cite{Fe30_Plateau}.
In addition, this cluster manifests a very broad Inelastic Neutron Scattering (INS) response 
as shown by Garlea \textit{et al.}~\cite{Garlea_INS}.
On the other hand, Mo$_{72}$V$_{30}$ has a much stronger AFM exchange $J/k_B\simeq 250$ K~\cite{V30_1,V30_2},
and thus is not well suited for the observation of the field-induced plateau. However, its
low-energy excitation spectrum can still be investigated by INS experiments
(which, to our knowledge, have not been performed so far).
As to the $s=1/2$ cuboctahedron Cu$_{12}$La$_8$~\cite{Cu12}, we are not aware of any magnetic
measurements reported so far on this cluster.

The main magnetic properties of the present clusters can be explained very well by
the isotropic Heisenberg model with a single AFM exchange parameter $J$, i.e.
\be\label{Heis.Eq}
\mc{H}= J \sum_{\langle ij\rangle} \vec{s}_i\cdot \vec{s}_j~,
\ee
where, as usual, $\langle i j \rangle$ denotes pairs of mutually interacting spins $s$ at sites $i$ and $j$.
Other terms such as single-ion anisotropy (for $s>1/2$) or Dzyaloshinsky-Moriya interactions
must be present as well in the present clusters, but they are expected to be much smaller than the exchange interactions 
and thus they can be neglected.
Here, as a simple theoretical tool to understand some of the properties of the Heisenberg model,
it will be very expedient to introduce some fictitious exchange anisotropy, i.e. extend Eq.~(\ref{Heis.Eq})
to its more general XXZ variant
\begin{eqnarray}
\mc{H}' &=& \mc{H}_z+\mc{H}_{xy},\label{XXZ.Eq}\\
\mc{H}_z &=& J_z \sum_{\langle ij\rangle} s_i^zs_j^z~,\label{Hz.Eq}\\
\mc{H}_{xy}&=& \frac{J_{xy}}{2}\sum_{\langle ij\rangle}(s_i^+s_j^-+s_i^-s_j^+)\label{Hxy.Eq}~,
\end{eqnarray}
where $J_{xy}$, $J_z$ denote the transverse and longitudinal exchange parameters respectively.
In what follows we denote $\alpha=J_{xy}/J_z$.

The main results presented in this article are of direct relevance to the experimental
findings in Mo$_{72}$Fe$_{30}$ mentioned above and thus span two major themes.
The first deals with the nature of the low-lying excitations above the $M=M_s/3$ plateau phase.
For the $s=1/2$ icosidodecahedron we show that all
these excitations are adiabatically connected to collinear ``up-up-down'' (henceforth ``uud'')
Ising ground states (GS's), at the same time being well isolated from higher levels by a relatively large energy gap.
We argue that this feature must be special to the topology of the icosidodecahedron and that it must
survive for $s=5/2$ as well. This prediction can be verified experimentally by a measurement of the
low-temperature specific heat and the associated entropy content at the plateau phase of Mo$_{72}$Fe$_{30}$.
A complementary physical picture will emerge by performing a high order perturbative
expansion in $\alpha$, in the spirit of Refs.~\onlinecite{Cabra,Bergman1,Bergman2}, and by deriving and solving
to lowest order the corresponding effective QDM on the dual clusters. The dependence of the model
parameters on $\alpha$ and $s$ is also found and given explicitly.

Our second theme concerns the origin of the broad INS response reported for Mo$_{72}$Fe$_{30}$~\cite{Garlea_INS}. 
Previous theories based on the excitations of the rotational band model~\cite{Garlea_INS,SchnackLubanModler} 
or on spin wave calculations~\cite{Cepas_LSW,Waldmann_LSW} predict a small number of discrete excitation lines at
low temperatures and thus cannot explain the broad INS features. Our interpretation of this behavior is based on the 
notion of the simultaneous presence of several rotational bands or towers of states at low energies 
which originate from the large degree of classical degeneracy, a generic feature of highly frustrated systems.
Indeed, our exact diagonalizations demonstrate the existence of an unusually high density of low-energy
excitations manifesting in the full magnetization range. A detailed group theory analysis reveals that the low-energy 
spectra are of semiclassical origin up to a relatively large energy cutoff.
We will also show that the symmetry of the corresponding excitations for $s=1/2$ does not conform with
this semiclassical picture.

A quite appealing feature of these molecular clusters is their high point group symmetry, namely the
full Octahedral group $\mathsf{O}_h=\mathsf{O}\times \mathsf{i}$ (with 48 elements) and the full Icosahedral
group $\mathsf{I}_h=\mathsf{I}\times \mathsf{i}$ (with 120 elements) for the cuboctahedron and the
icosidodecahedron respectively (here $\mathsf{i}$ denotes the inversion).
This allows for a drastic reduction of the dimensionality of the problem.
In order to fully exploit all symmetry operations we have employed a generalization~\cite{nikos}
of the standard ED technique~\cite{ED} so as to treat higher than one-dimensional
Irreducible Representations (IR's) also. With this approach, one is able to classify the energy levels
according to both $S_z$ and the IR of the point group while resolving their full degeneracy.

The remainder of the article is organized as follows. In Sec.~\ref{Col_vs_Kagome.Sec} we discuss some of the
spectral features of the present kagom\'e-like nanomagnets (with particular emphasis on localized magnons)
and contrast them to typical spectra of unfrustrated AFM's. This is illustrated by comparing with a simple 12-site 
bipartite $s=1/2$ cluster. The investigation of the nature of the $M=M_s/3$ plateau is presented in Sec.~\ref{Plateau.Sec}.
This includes both analytical and numerical results from high order degenerate perturbation theory around the Ising limit, 
the construction of the associated effective QDM's and their extrapolation to the Heisenberg limit. 
We also discuss the case of higher $s$ and the connection to the plateau phase of Mo$_{72}$Fe$_{30}$.
In Sec.~\ref{Coplanarity.Sec} we demonstrate the presence of several low-energy rotational bands in $s>1/2$ Heisenberg 
spectra and reveal their semiclassical origin. The analysis is based on a careful comparison to the symmetry properties 
of the semiclassical states and follows the basic lines of the seminal works of Bernu \textit{et al.}~\cite{Bernu1,Bernu2} 
and Lecheminant \textit{et al.}~\cite{Lecheminant_J1J2,Lecheminant_kagome} on this subject in the context of the 
triangular and kagom\'e AFM. Predictions for the corresponding tower of states are also given for the ($s>1/2$) icosidodecahedron.
Our core idea of the presence of several rotational bands due to the large spatial degeneracy of the classical states 
is also exemplified in Sec.~\ref{XY.Subsec} by a discussion of the much simpler case of the $s>1/2$ XY model.
We leave Sec.~\ref{Discussion.Sec} for an overview of the major findings of this work.
In order for the manuscript to be self-contained, we include two appendices.
In Appendix~\ref{DPT.App} we summarize the main aspects of the degenerate perturbation expansion
around the Ising limit, while in Appendix~\ref{Tower.App} we give the details of the derivation
of the full symmetry properties of the semiclassical towers of states for the Heisenberg and the XY model.

A special remark is in order here regarding our choice of presentation of the spectra.
Since we are interested in the low-energy excitations in the whole magnetization range (these are the
most accessible and thus most relevant as one ramps up an external field at low temperatures)
and in order to best illustrate the central features, we have chosen to
(except for Fig.~\ref{12mer_Tower.Fig}) shift the lowest energy $E_0(S_z)$ (or $E_0(S)$) of each $S_z$ ($S$) sector to zero.
This guarantees a fine resolution of the low-energy spectra in the whole magnetization range.

\begin{figure}%[!b]
\centering
\includegraphics*[width=0.35\linewidth, angle=90]{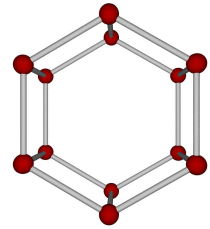}
\caption{(Color online) Schematic representation of the unfrustrated, bipartite 12-site cluster
discussed in Sec.~\ref{Col_vs_Kagome.Sec}. Its symmetry group is $\mathsf{D}_{6h}=\mathsf{D}_6\times \mathsf{i}$.}
\label{12mer.Fig}
\end{figure}

\section{Unfrustrated vs. kagom\'e-like AFM's : General spectral features (s=1/2)}\label{Col_vs_Kagome.Sec}
\begin{figure}%[!t]
\centering
\includegraphics*[width=0.9\linewidth]{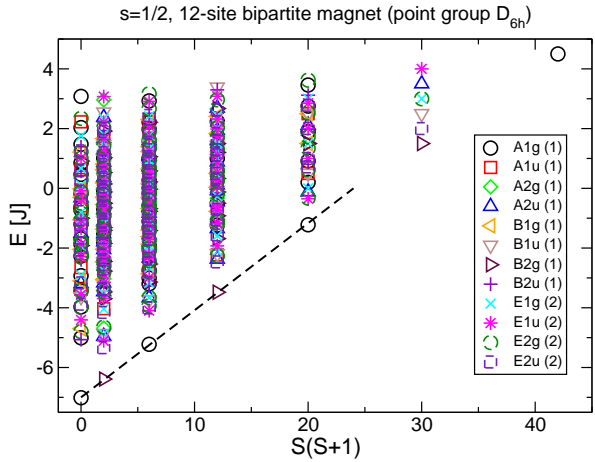}
\caption{(Color online) Low-energy spectrum of the 12-site bipartite AFM shown in Fig.~\ref{12mer.Fig}
as a function of $S(S+1)$ and classified according to IR's of the $\mathsf{D}_{6h}$ group.
The dotted line denotes the Anderson tower of states which embodies the finite-size features of the
spatial and $\mathsf{SU}(2)$ broken N\'eel state in the thermodynamic limit.}
\label{12mer_Tower.Fig}
\end{figure}

Our main purpose in this section is to present the low-energy spectra of the $s=1/2$ Heisenberg
cuboctahedron and icosidodecahedron and to highlight their main features which are common in all frustrated
AFM's. For comparison, it is expedient to also present the energy spectrum of a bipartite unfrustrated
magnet. To this end, we have chosen the hypothetical 12-site cluster depicted in Fig.~\ref{12mer.Fig}.
The symmetry of this cluster is the dihedral group $\mathsf{D}_{6h}=\mathsf{D}_6\times \mathsf{i}$
which consists of 24 elements. 
Figure~\ref{12mer_Tower.Fig} shows the low-energy Heisenberg spectrum as a
function of $S(S+1)$, classified according to the 12 different IR's of $\mathsf{D}_{6h}$ 
(cf. Ref.~\onlinecite{GroupTheory}) shown in the legend.
The spectrum is typical of unfrustrated AFM's~\cite{Lhuillier_Towers,Richter_chapter} of finite size $N$.
For instance, we may associate the lowest energy band indicated by the dotted line in
Fig.~\ref{12mer_Tower.Fig} with the so-called Anderson tower of states~\cite{Lhuillier_Towers},
which is the finite-size manifestation of the $\mathsf{SU}(2)$ symmetry breaking process occurring in the
thermodynamic limit.
As can be seen in Fig.~\ref{12mer_Tower.Fig}, this tower consists entirely of the two one-dimensional
representations A1g and B2g of $\mathsf{D}_{6h}$, which alternate between even and odd $S$ respectively.
The physics behind this symmetry structure is intimately related to the symmetry properties of the
semiclassical two-sublattice N\'eel state. For instance, the combinations A1g $\pm$ B2g transform into each other in
exactly the same way as the two spatial counterparts of the N\'eel state. Above the lowest tower of states of
Fig.~\ref{12mer_Tower.Fig} there exists a finite excitation gap followed by a quasi continuum of higher excitations.
All these features are typical of unfrustrated AFM's.
%%%%%%%%%%%%%%%%%%%%%%%%%%%%%
%%%%%%%%%%%%%%%%%%%%%%%%%%%%%

%%%%%%%%%%%%%%%%%%%%%%%%%%%%%
\begin{figure}
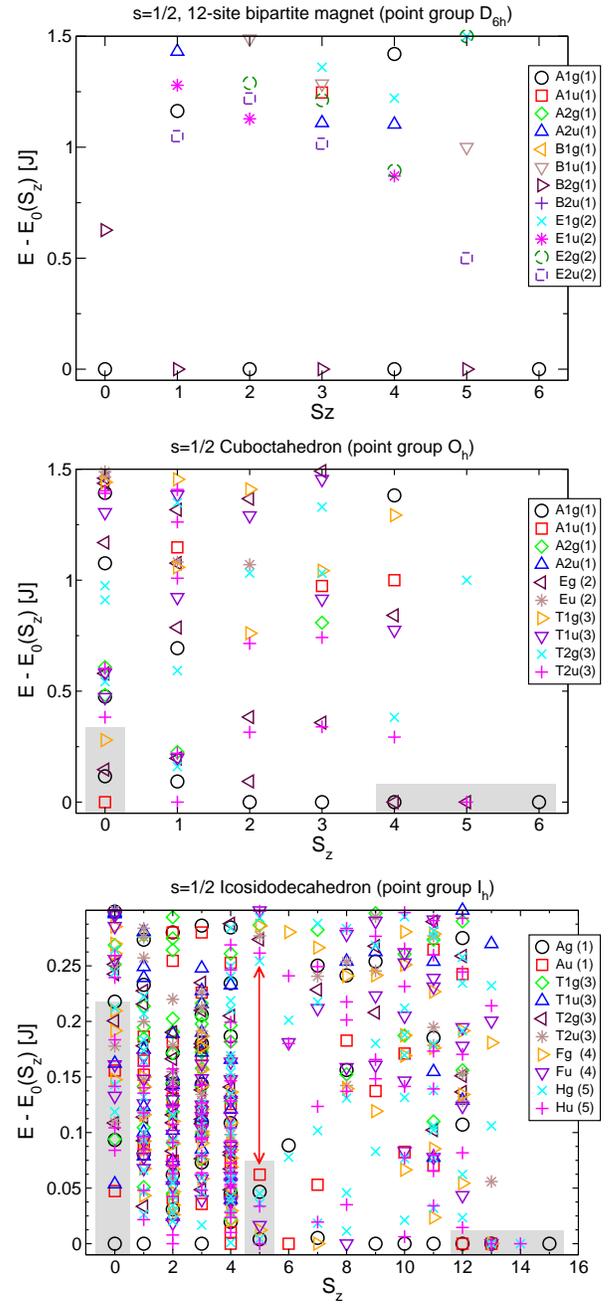
%[!t]
\centering
\includegraphics*[width=0.9\linewidth]{12merOneHalf_Heis_Zoom}\\
\vspace{2mm}
\includegraphics*[width=0.9\linewidth]{CubocOneHalf_Heis_Zoom}\\
\vspace{2mm}
\includegraphics*[width=0.9\linewidth]{IcosiOneHalf_Heis}
\caption{(Color online) Low-energy spectra (shifted as described in the text) of the $s=1/2$ Heisenberg model
on the 12-site unfrustrated magnet shown in Fig.~\ref{12mer.Fig} (top), on the cuboctahedron (middle)
and the icosidodecahedron (bottom). Three special features are highlighted by the corresponding shaded areas
in the two lower panels:
(i) the low-lying singlets below the first triplet in the $S_z=0$ sectors,
(ii) the existence of localized magnons highlighted in the sectors below saturation, and
(iii) the lowest 36 Ising-like configurations (cf.Sec.~\ref{Icosi_Plateau.Subsubsec}) above the
plateau $Sz=5$ sector of the icosidodecahedron case (lowest panel).
The large energy gap between these configurations and higher excitations is indicated by the arrow.}
\label{12mer_Cuboc_Icosi_Heis.Fig}
\end{figure}
%%%%%%%%%%%%%%%%%%%%%%%%%%%%%

In contrast, frustrated AFM's show very different low-energy features as exemplified by the $s=1/2$ spectra
shown in the two lower panels of Fig.~\ref{12mer_Cuboc_Icosi_Heis.Fig} for the two molecular magnets of
the present study. For comparison, the upper panel shows the low-energy portion of Fig.~\ref{12mer_Tower.Fig} in terms of $S_z$.
The contrast between the two types of spectra is more than evident (note that both of
the upper two panels correspond to 12-site clusters and are shown in the same energy scale).
The most striking feature emerging in frustrated AFM's is the absence of a clear energy scale separating
a lowest band from higher lying excitations.
Instead, a large ``bulk'' of low-energy excitations is manifested in the whole range of $S_z$ forming a
quasi-continuum. This is a central feature that holds also for higher $s$ (cf. Sec.~\ref{Coplanarity.Sec}) 
and stems from the highly frustrated exchange interactions in these clusters.
The nature of these excitations for $s=1/2$ is not completely understood~\cite{Mila}
but, as we are going to show in Sec.~\ref{Coplanarity.Sec},
a qualitative understanding can be obtained for $s>1/2$ based on the large classical degeneracy of
spin configurations which remains dominant in the semiclassical regime.
In particular, the broad INS response reported in Ref.~\onlinecite{Garlea_INS} for Mo$_{72}$Fe$_{30}$
is naturally accounted for by the results of this analysis.

%%%%%%%%%%%%%%%%%%%%%%%%%%%%%
\begin{table}%[!b]
\caption{Lowest energies $E_0(S_z)$ of each $S_z$ sector and the corresponding degeneracies
for the $s=1/2$ cuboctahedron (a) and icosidodecahedron (b).}
\label{LowestEnergies.Table}

\begin{ruledtabular}
\vspace{0.1cm}\raggedright (a) $\mathbf{s=1/2}$ \textbf{cuboctahedron}
\begin{tabular}{ccc|ccc}
$S_z$ & $E_0(S_z) [J]$ & deg. & $S_z$ & $E_0(S_z) [J]$ & deg.\\
\hline
0 & -5.44487521 &1  &4& 0&3 \\
1 & -5.06220685 &3  &5& 3&5 \\
2 & -4.36867379 &1  &6& 6&1 \\
3 & -2.63135381 &1
\end{tabular}
\end{ruledtabular}

\begin{ruledtabular}
\vspace{0.3cm}\raggedright (b) $\mathbf{s=1/2}$ \textbf{icosidodecahedron}
\begin{tabular}{ccc|ccc}
$S_z$ & $E_0(S_z) [J]$ & deg. & $S_z$ & $E_0(S_z) [J]$ & deg.\\
\hline
0 & -13.23421620&1   & 8 & -4.80706643&4\\
1 & -13.01640033&1   & 9 & -2.41759676&1\\
2 & -12.61867043&5   & 10&  0.31845649&1\\
3 & -12.05650773&1   & 11&  3.12078845&1\\
4 & -11.22383327&1   & 12&  6&2\\
5 & -10.30278977&5   & 13&  9&25\\
6 &  -8.95866550&1   & 14& 12&10\\
7 &  -7.01225008&4   & 15& 15&1
\end{tabular}
\end{ruledtabular}

\end{table}
%%%%%%%%%%%%%%%%%%%%%%%%%%%%%
%%%%%%%%%%%%%%%%%%%%%%%%%%%%%
\begin{figure}[!t]
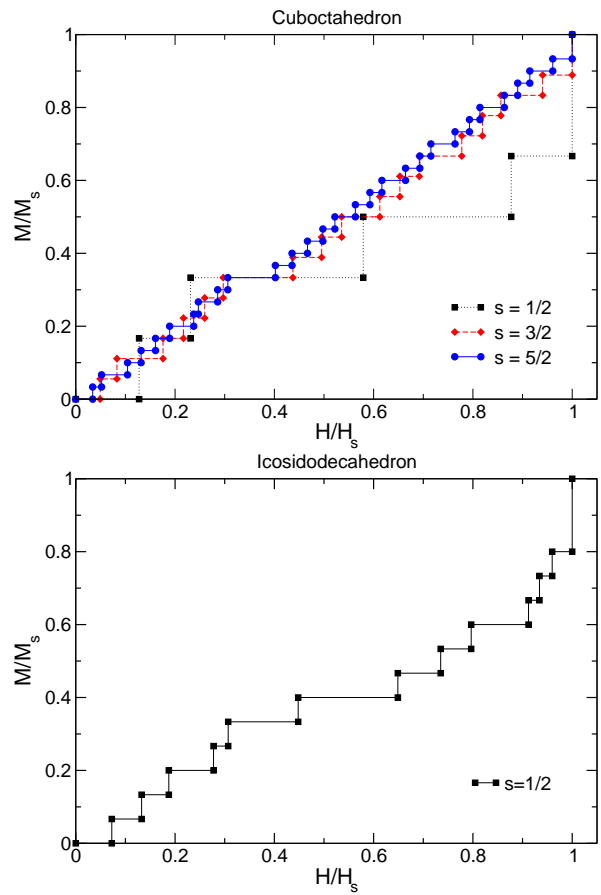

\centering
\includegraphics*[width=0.9\linewidth]{MvsB_Cuboc}\\%\vspace{0.14in}
\includegraphics*[width=0.9\linewidth]{MvsB_Icosi}
\caption{(Color online) Zero-temperature magnetization curves for the Heisenberg cuboctahedron
($s=1/2,3/2,5/2$) and the $s=1/2$ icosidodecahedron
(see also Refs.~\onlinecite{Schmidt_Singlets,Schnack_ind_magn_MMs}).
The saturation magnetization and field values are given by $M_s= N_s s (g\mu_B)$
(where $N_s$ is the number of sites) and $H_s=6 s J/(g\mu_B)$.}
\label{MvsB.Fig}
\end{figure}
%%%%%%%%%%%%%%%%%%%%%%%%%%%%%

Let us now describe shortly some special spectral features and their origin. The ground state energies $E_0(S_z)$ for the 
two nanomagnets for $s=1/2$ are given in Table~\ref{LowestEnergies.Table}. These energies determine the zero-temperature
magnetization processes shown in Fig.~\ref{MvsB.Fig}. For the excitations above the ground state, three regimes of special 
interest can be highlighted (see shaded areas in Fig.~\ref{12mer_Cuboc_Icosi_Heis.Fig}):
(i) the singlet excitations below the lowest triplet\footnote{The presence of these singlets has been revealed previously 
by the work of R. Schmidt \textit{et al.} in Ref.~\onlinecite{Schmidt_Singlets} but the exact numbers and degeneracies could not be 
resolved by the reduced symmetry ED method used there.} which are given in Table~\ref{Singlets.Table} and amount to 7 for the cuboctahedron and 
80 for the icosidodecahedron\footnote{Quite interestingly, this number equals approximately $1.15727 ^{30}$ which is quite 
close to the reported~\cite{Waldtmann} scaling of $\approx 1.15^N$ for the number of low-lying singlets in the kagom\'e 
(with even $N$) despite the fact that the icosidodecahedron contains pentagons and not hexagons.}, 
(ii) the existence of degenerate localized magnons below saturation and
(iii) the presence of a number of well isolated low-lying states right above the $M=M_s/3$ ($S_z=5$) plateau phase 
of the $s=1/2$ icosidodecahedron. The latter will be analyzed in detail in Sec.~\ref{Plateau.Sec}.

%%%%%%%%%%%%%%%%%%%%%%%%%%%%%
\begin{table}[!t]
\caption{(a) Energies of the seven lowest singlets of the $s=1/2$ Heisenberg cuboctahedron
lying below the first triplet $E=-5.06220685 J$ (T2u) state, together with their $\mathsf{O}_h$
symmetry classification and their degeneracy.
(b) Energies of the 80 lowest singlets of the $s=1/2$ Heisenberg icosidodecahedron lying below the first
triplet $E=-13.01640033 J$ (Ag), together with their $\mathsf{I}_h$ classification.}
\label{Singlets.Table}
\begin{ruledtabular}
\vspace{0.1cm}\raggedright (a) $\mathbf{s=1/2}$ \textbf{cuboctahedron}
\begin{tabular}{cc|cc}
Energy [J] & IR (deg) & Energy [J] & IR (deg) \\
\hline
-5.44487521 & A1u(1) & -5.29823654 & Eg (2)\\
-5.32839240 & A1g(1) & -5.16529346 & T1g(3)\\
\end{tabular}
\end{ruledtabular}

\begin{ruledtabular}
\vspace{0.3cm}\raggedright (b) $\mathbf{s=1/2}$ \textbf{icosidodecahedron}
\begin{tabular}{cc|cc}
Energy [J] & IR (deg) & Energy [J] & IR (deg) \\
\hline
-13.23421620 & Ag (1) & -13.09125447 & Hg (5)\\
-13.18689258 & Au (1) & -13.08659708 & Fg (4)\\
-13.18057238 & T1u(3) & -13.07844898 & Au (1)\\
-13.15013156 & Hu (5) & -13.07310588 & Fu (4)\\
-13.14089964 & Ag (1) & -13.07200565 & T1u(3)\\
-13.14024171 & T1g(3) & -13.05645698 & T2u(3)\\
-13.12997109 & Hu (5) & -13.05072896 & Hu (5)\\
-13.12560855 & T2g(3) & -13.04261651 & Fg (4)\\
-13.12514475 & T2u(3) & -13.03366847 & T2g(3)\\
-13.11552338 & Hg (5) & -13.02470946 & Fg (4)\\
-13.10136600 & Fu (4) & -13.02203094 & Hg (5)\\
-13.09264778 & Hu (5) &
\end{tabular}
\end{ruledtabular}
\end{table}
%%%%%%%%%%%%%%%%%%%%%%%%%%%%%

The concept of localized magnons has been largely discussed in the context of highly frustrated
bulk AFM's~\cite{Richter_chapter,Schulenburg_ind_magn,Schmidt_ind_magn,Derzhko_ind_magn,Schnack_ind_magn_bulk,
Richter_JPhys,Mike_Caloric}. For the present clusters it has been discussed by Schnack \textit{et al.}~\cite{Schnack_ind_magn_MMs,Schnack_caloric}.
We shortly revisit this issue here in the light of our symmetry resolved method.
Quite generally, the eigenvalues of $\mc{H}$ in the one magnon ($M=M_s-1$) subspace are equal
(apart from an overall constant energy shift) to the eigenvalues of the adjacency matrix
$(\mc{J}_{\mu\nu})$ times the spin $s$~\cite{SchmidtLuban}.
Our decomposition of the respective subspaces for the cuboctahedron and the icosidodecahedron
in terms of IR's of the full $\mathsf{O}_h$ and $\mathsf{I}_h$ groups are
(in order of increasing energy): (Eg$\oplus$T2u)$\oplus$T2g$\oplus$T1u$\oplus$A1g, and
(Hg$\oplus$Hu)$\oplus$T2u$\oplus$Fg$\oplus$Fu$\oplus$Hg$\oplus$T1u$\oplus$Ag respectively, and they
are compatible to the ones given in Ref.~\onlinecite{SchmidtLuban} in terms of IR's of the $\mathsf{O}$
and $\mathsf{I}$ subgroups.

%%%%%%%%%%%%%%%%%%%%%%%%%%%%%
\begin{figure}%[!t]
\centering
\includegraphics*[width=0.8\linewidth]{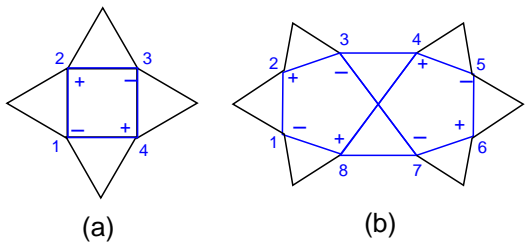}
\caption{(Color online) The minimal loops hosting the localized $k=\pi$ magnons on the topology of the
cuboctahedron (a) and the icosidodecahedron (b). They read
$|\psi\rangle = \frac{1}{2}\left(s_1^--s_2^-+s_3^--s_4^-\right)|0\rangle$, and
$|\psi\rangle = \frac{1}{2\sqrt{2}} \left(s_1^--s_2^-+\ldots-s_8^-\right)|0\rangle$ respectively,
where $|0\rangle$ is the ferromagnetic vacuum. These are exact eigenstates of Eq.~(\ref{Heis.Eq}) owing to
a cancellation of interaction terms resulting from the special corner-sharing triangles
topology~\cite{Schulenburg_ind_magn}.
In both cases, the energy $\hbar\omega_m$ required to excite these localized states
measured from the ferromagnetic (FM) vacuum $|0\rangle$ ($E_0=N_b J s^2$, where $N_b$ is the number of bonds)
equals $\hbar\omega_m =- 6 s J$, and is independent of the length of the loops.}
\label{MagnonLoops.Fig}
\end{figure}
%%%%%%%%%%%%%%%%%%%%%%%%%%%%%

For the cuboctahedron, the lowest one-magnon level is 5-fold degenerate (Eg$\oplus$T2u),
see Fig.~\ref{12mer_Cuboc_Icosi_Heis.Fig}(middle). These correspond to localized, non-interacting magnon
states. The smallest loops that can host such magnons are the square faces depicted in
Fig.~\ref{MagnonLoops.Fig}(a). The 5-fold degeneracy is due to the fact that there are 6 different
square faces on this cluster, but not all magnons are independent: The sum of all 6 square magnons taken
with opposite phases in neighboring squares vanishes.
The lowest energy level of the $S_z=4$ two-magnon space is 3-fold degenerate (Ag$\oplus$Eg), and corresponds
to the 3 different ways of placing two non-interacting magnon excitations (there are three different pairs
of opposite squares). Placing one more magnon gives an interaction energy cost and a
non-degenerate $S_z=3$ lowest level. We should remark here that magnon states ``living'' on the
hexagonal equators of the cluster are also exact eigenstates, but each of these can be easily expressed
as a linear combination of surrounding square magnons.
The lowest level degeneracies for the one-magnon and the two-magnon space
are in agreement with the values of $N/3+1$ and $N^2/18-N/2+1$ respectively with $N=12$ which are derived
in Ref.~\onlinecite{Derzhko_ind_magn} for the kagom\'e lattice (for which the hexagonal loops are the most
local and natural ones for the description of the localized magnons).

For the icosidodecahedron, the lowest one-magnon level is 10-fold degenerate (Hu$\oplus$Hg).
Here, the smallest loops that can host such localized states are the octagons
surrounding a given vertex and depicted in Fig.~\ref{MagnonLoops.Fig}(b).
The 10-fold degeneracy can be attributed to a non-trivial linear dependence among
the 30 different octagonal magnons on this cluster.
The lowest energy level of the two-magnon manifold is 25-fold degenerate and decomposes into
$\text{Ag} \oplus \text{Au} \oplus \text{Fg} \oplus \text{Fu} \oplus 2\text{Hg} \oplus \text{Hu}$.
Hence, there exist 25 ways of placing two mutually non-interacting magnons. Similarly, the lowest energy
of the three-magnon space is two-fold degenerate (Ag$\oplus$Au), whereas that of the $S_z=11$ sector
is non-degenerate, signifying that it is not possible to have four magnons without an interaction energy cost.

The existence of localized, non-interacting magnon states results in a magnetization jump of $\Delta S_z >1$,
since the lowest energies at the corresponding $S_z$ sectors scale linearly with the number of
magnons, and thus cross each other at the same (saturation) field.
We remark here that all features related to the existence of localized magnons
(symmetry decomposition, degeneracy, and the magnetization jump in absolute units)
do not depend on the value of $s$ (see e.g. Fig.~\ref{Cuboc_Heis.Fig} below).
Finally, the fact that the number of independent magnons is larger in the icosidodecahedron
than the cuboctahedron case is clearly related to their size.
In extended frustrated AFM's, this number grows exponentially with system size
but depends in a non-trivial way on the topology of the system and is connected to
the question of linear independence~\cite{Schmidt_ind_magn,Derzhko_ind_magn}.
The extensive degeneracy gives rise to a macroscopic magnetization jump at the saturation field
and a large magnetocaloric effect (see e.g. Ref.~\onlinecite{Mike_Caloric}).
A study of the latter on the present clusters can be found in Ref.~\onlinecite{Schnack_caloric}.

\section{$M=M_s/3$ Plateau phase}\label{Plateau.Sec}
In this section, we focus on the nature of the excitations above the $M=M_s/3$ plateau.
There are two major reasons for paying special attention to this particular plateau
among the remaining ones which are present anyway in our finite-size clusters (cf.Fig.~\ref{MvsB.Fig}).
The first is of practical interest and is related to the experimental manifestation~\cite{Fe30_Plateau}
of this particular phase in the $s=5/2$ Mo$_{72}$Fe$_{30}$ cluster.
Besides, as shown in the upper panel of Fig.~\ref{MvsB.Fig}, the $M=M_s/3$ plateau seems to be the
most stable and survives at finite $s>1/2$ (the staircase $s=1/2$ magnetization process eventually turns
into the expected (classical) linear behavior~\cite{Fe30_Plateau} for very large $s$).
The second reason is that the $M=M_s/3$ plateau phase is a generic feature of frustration and is known
to survive in the thermodynamic limit for some bulk AFM's (cf. Ref.~\onlinecite{Richter_chapter}).

In Sec.~\ref{Thermo.Subsec} we present and analyze a striking feature of the excitations above the 
plateau phase of the $s=1/2$ icosidodecahedron and show how it can be observed experimentally in thermodynamic 
measurements. In Sec.~\ref{QDM.Subsec} we present our derivation of an effective Quantum Dimer Model for the plateau 
and reveal the major role of the topology and the spin $s$.

%%%%%%%%%%%%%%%%%%%%%%%%%%%%%
\begin{table}
\caption{Energies (in units of $J$) of the 36 lowest $S_z=5$ states of the $s=1/2$ Heisenberg
icosidodecahedron together with their $\mathsf{I}_h$ classification. The lowest excitation above this
manifold lies at $E=-10.04843786$ (Hg).}
\label{LowestSz5Icosi.Table}
\begin{ruledtabular}
\begin{tabular}{cc|cc}
 Energy [J]& IR(deg) & Energy [J]& IR(deg)\\
\hline
-10.30278978 & Hu (5)   & -10.26904953  & Hu (5)\\
-10.29875816 & Ag (1)   & -10.26657194  & Hg (5)\\
-10.29837409 & Hg (5)   & -10.25765943  & Hg (5)\\
-10.29057364 & Fg (4)   & -10.25604215  & Ag (1)\\
-10.28622445 & Fu (4)   & -10.24060604  & Au (1)
\end{tabular}
\end{ruledtabular}
\end{table}
%%%%%%%%%%%%%%%%%%%%%%%%%%%%%
%%%%%%%%%%%%%%%%%%%%%%%%%%%%%
\begin{figure}%[!b]
\centering
\includegraphics*[width=0.9\linewidth]{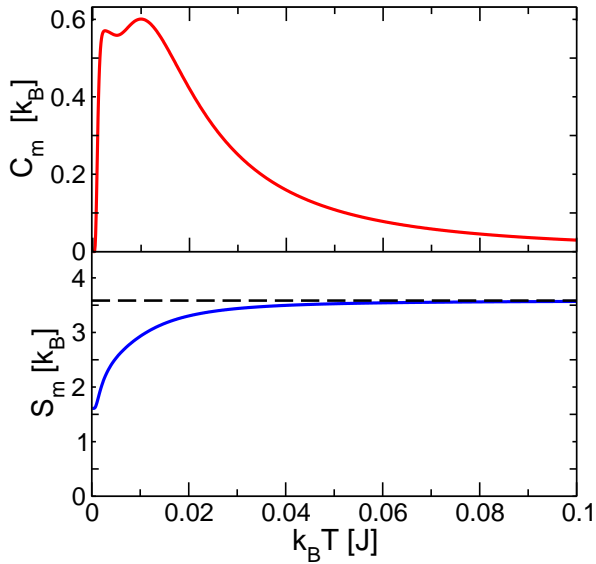}
\caption{(Color online) Temperature dependence of the magnetic entropy $S_m$ and specific heat $C_m$ 
(in units of Boltzmann's constant $k_B$) content of the lowest 36 plateau states of 
$s=1/2$ Heisenberg icosidodecahedron. Here the magnetic field corresponds to the
center of the $M=M_s/3$ plateau.
The entropy starts off from the value $\ln{5}\simeq 1.6094$ corresponding to the lowest
5-fold degenerate Hu state. The dashed line denotes the entropy value
of $\ln{36}\simeq 3.5835$ corresponding to the full 36-fold low-energy subspace.}
\label{EntropySpecHeat.Fig}
\end{figure}
%%%%%%%%%%%%%%%%%%%%%%%%%%%%%

\subsection{Thermodynamics}\label{Thermo.Subsec}
Our Exact Diagonalizations for the $s=1/2$ icosidodecahedron shown in the lowest panel
of Fig.~\ref{12mer_Cuboc_Icosi_Heis.Fig} reveal a striking feature at the $M=M_s/3$ sector:  
The low-lying excitation spectrum immediately above the plateau
consists of a group of 36 states (their energies are given in Table~\ref{LowestSz5Icosi.Table})
which are well isolated from higher excitations by a gap of order $0.2 J$, an order of magnitude
larger than the excitations ($\sim 0.01 J$) within this manifold.
We argue below that this peculiar feature must be related to the special topology of the icosidodecahedron
(it does not appear for the cuboctahedron).
Given this behavior, it is expedient to consider the low-temperature dependence of the magnetic specific
heat and entropy content of the lowest 36 plateau states. These quantities are shown in
Fig.~\ref{EntropySpecHeat.Fig} for the $s=1/2$ case and for $g\mu_B H \simeq 1.1326 J$
(where $g$ is the electronic spectroscopic factor and $\mu_B$ the Bohr magneton) which corresponds
to the center of the $M=M_s/3$ plateau. At this field value, the lowest excitations of the adjacent
$S_z=4$ and $S_z=6$ sectors lie approximately $0.2 J$ above the ground manifold
(higher $S_z=5$ states lie $\sim 0.25 J$ above).
Hence, the temperature behavior shown in Fig.~\ref{EntropySpecHeat.Fig} must be valid at
$k_B T \lesssim 0.1 J$. In this temperature regime, the entropy content of the lowest $S_z=5$ states
is already saturated to its full value of $\ln{36}\simeq 3.5835$ (dashed line) which amount to a
sizable fraction of about $17 \%$ of the full $30 \times \ln{2} \simeq 20.7944$ magnetic entropy
of the cluster. The fine details of Fig.~\ref{EntropySpecHeat.Fig} can be associated to the actual splitting
between the 36 states. For instance, the double-peak form of the specific heat stems from
the small separation of the first 19 from the remaining 17 states (cf.Table~\ref{LowestSz5Icosi.Table}
and lowest panel of Fig.~\ref{12mer_Cuboc_Icosi_Heis.Fig}). Note also that the entropy starts off from the value 
$\ln{5}\simeq 1.6094$ corresponding to the lowest 5-fold degenerate Hu state (cf.Table~\ref{LowestSz5Icosi.Table}).

Unfortunately the plateau regime of Mo$_{72}$V$_{30}$ cannot be reached experimentally due to the large exchange 
value of $J/k_B\simeq 250$ K. We shall argue below, based on the results of our effective QDM, that a similar structure 
must exist above the $M=M_s/3$ plateau of the $s=5/2$ Mo$_{72}$Fe$_{30}$ cluster. In particular, we shall argue that
(i) the lowest 36 states are split into two almost degenerate levels of 30 and 6 states respectively 
with the former being lowest in energy, and (ii) that a sizable gap between these 36 states and higher excitations 
must probably survive as well. The corresponding specific heat peak can be verified by thermodynamic measurements 
on the Mo$_{72}$Fe$_{30}$ cluster at the plateau regime of $H\simeq 5.9$ Tesla. Despite the very small exchange value 
($J\simeq 1.57$ K) of Mo$_{72}$Fe$_{30}$ (one presumably needs to reach ultra low temperatures, $T\lesssim 200$ mK) one 
may still confirm our picture by an assessment of the missing entropy~\cite{Ramirez,Bramwell}.

%%%%%%%%%%%%%%%%%%%%%%%%%%%%%
\begin{figure}
\centering
\includegraphics*[width=0.7\linewidth]{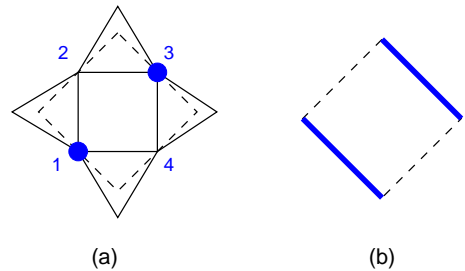}
\caption{(Color online) A local view of one of the ``uud'' Ising configurations (with two spins
pointing up and one down in \textit{each} triangle) and the mapping between vertices of the
cuboctahedron (solid line in (a)) and edges of its dual cluster, the cube (dashed lines).
In (a), all spins point up except the ones at vertices $1$ and $3$ (designated by the black dots) which point down. 
By mapping each down spin in (a) to a dimer on the corresponding edge of the cube we obtain the dimer plaquette in (b).
As discussed in Sec.~\ref{Cuboc_Plateau.Subsubsec}, such square loops with alternating up-down spins
have the minimum even length and thus govern the lowest order kinetic processes driven by $\mc{H}_{xy}$.
In (a), these read $t~ s_1^+s_2^-s_3^+s_4^-$ with $t \propto \alpha^{4s}$ and map to the dimer plaquette
flip of Eq.~(\ref{CubocQDM.Eq}). The lowest order diagonal processes are also confined on these square loops
and scale as $v\propto \alpha^4$.}\label{CubocOffDiagLoop.Fig}
\end{figure}
%%%%%%%%%%%%%%%%%%%%%%%%%%%%%
%%%%%%%%%%%%%%%%%%%%%%%%%%%%%
\begin{figure}
\centering
\includegraphics*[width=0.8\linewidth]{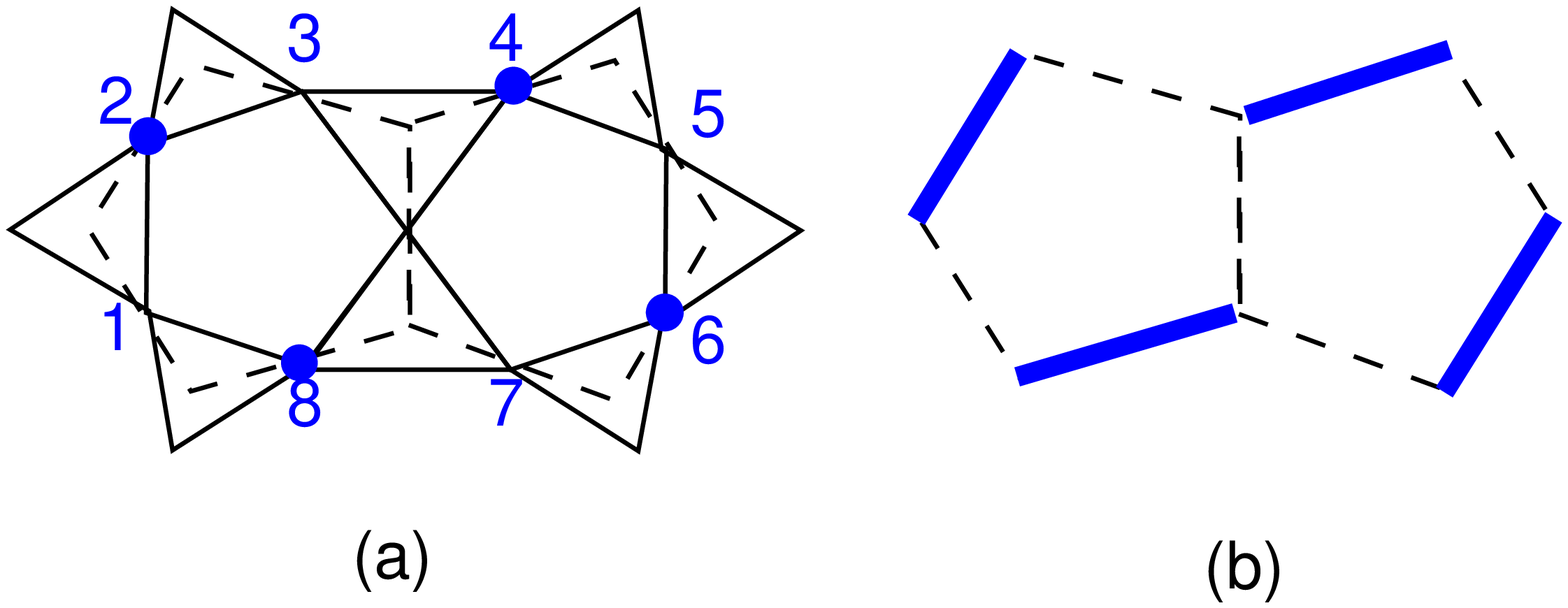}
\caption{(Color online) A local view of one of the ``uud'' Ising configurations (with two spins
pointing up and one down in \textit{each} triangle) and the mapping between vertices of the
icosidodecahedron (solid line in (a)) and edges of its dual cluster, the dodecahedron (dashed lines).
In (a), all spins point up except the ones at vertices $2$, $4$, $6$ and $8$
which point down. By mapping each down spin in (a) to a dimer on the corresponding edge of the dodecahedron
one obtains the dimer plaquette in (b). As discussed in Sec.~\ref{Icosi_Plateau.Subsubsec},
such octagonal loops with alternating up-down spins have the minimum even length
and thus govern the lowest order kinetic processes driven by $\mc{H}_{xy}$. In (a), these read
$t~ s_1^-s_2^+s_3^-\ldots s_8^+$ with $t\propto \alpha^{8s}$ and map to the dimer plaquette flips
of Eq.~(\ref{Icosi_Heff.Eq}). On the other hand, the lowest order diagonal processes
are confined in a single pentagon and scale as $v\propto \alpha^5$.}
\label{IcosiOffDiagLoop.Fig}
\end{figure}
%%%%%%%%%%%%%%%%%%%%%%%%%%%%%

\subsection{Effective QDM}\label{QDM.Subsec}
In what follows we present a complementary picture for the nature of the excitations above the plateau phase. 
This picture reveals the central role of the topology and the intrinsic spin $s$ and will emerge from the 
derivation of an effective quantum dimer model (QDM) in the spirit of Refs.~\onlinecite{Cabra,Bergman1,Bergman2} .
The main idea is to start from the degenerate $M/M_s=1/3$ ground state configurations of the Ising limit
and establish an adiabatic connection to the low-lying excitations of the Heisenberg point by employing
a perturbative expansion in the anisotropy parameter $\alpha=J_{xy}/J_z$. The resulting effective Hamiltonian
can be cast into the form of a QDM on the dual clusters as exemplified in Figs.~\ref{CubocOffDiagLoop.Fig}
and~\ref{IcosiOffDiagLoop.Fig} for the cuboctahedron and the icosidodecahedron respectively. 
For a general spin $s$, the $M=M_s/3$ ground state manifold of the Ising Hamiltonian $\mc{H}_z$
spans all configurations with two spins having $m=s$ and one with $m=-s$ \textit{in each triangle}.
Each one of these ``uud'' states on the cuboctahedron and the icosidodecahedron is in one-to-one
correspondence to a closed-packed dimer covering on their dual clusters, the cube and the dodecahedron respectively.

\subsubsection{Cuboctahedron}\label{Cuboc_Plateau.Subsubsec}
We start with the ``uud'' GS's of the Ising cuboctahedron. There are nine such states since this is the number 
of different dimer coverings on the cube. This manifold, henceforth $\mathbf{P}_\textrm{uud}$, decomposes into two invariant 
(under $\mathsf{O}_h$) ``uud'' families $\mathbf{P}^{(6)}_{\text{uud}}$ and $\mathbf{P}^{(3)}_{\text{uud}}$ with 6 and 3 
states respectively:
\be\label{CubocIsingRed.Eq}
\mathbf{P}_\textrm{uud} = \mathbf{P}^{(6)}_{\text{uud}}\oplus\mathbf{P}^{(3)}_{\text{uud}}~,
\ee
where
\be
\mathbf{P}^{(6)}_{\text{uud}} = \textrm{A1g}\oplus\textrm{Eg}\oplus\textrm{T2u},~
\mathbf{P}^{(3)}_{\text{uud}} = \textrm{A1g}\oplus\textrm{Eg}~.
\ee
Each of the two families contains states with a fixed number $n_c$ (2 and 4 respectively) of square plaquettes of the type 
of Fig.~\ref{DiagSquare.Fig}(c) (this number remains invariant under the operations $\mathsf{O}_h$ of the 
cluster). 

%%%%%%%%%%%%%%%%%%%%%%%%%%%%%
\begin{figure}
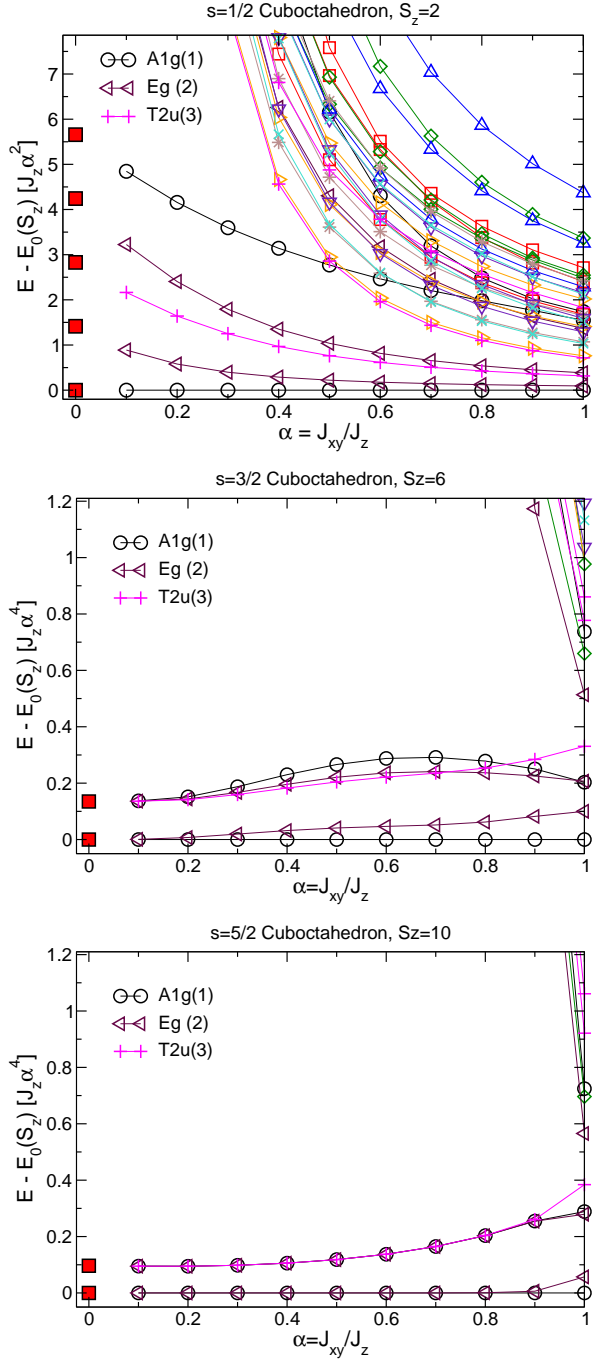
%[!b]
\centering
\includegraphics*[width=0.9\linewidth]{CubocOneHalf_XXZ}\\
\vspace{2mm}
\includegraphics*[width=0.9\linewidth]{CubocThreeHalf_XXZ}\\
\vspace{2mm}
\includegraphics*[width=0.9\linewidth]{CubocFiveHalf_XXZ}
\caption{(Color online) Lowest energy spectrum of the XXZ model on the cuboctahedron as a function of
$\alpha=J_{xy}/J_z$, for $s=1/2$ (top), $3/2$ (middle), and $s=5/2$ (bottom).
The energies are given in units of $J_z\alpha^2$ (top) and $J_z\alpha^4$ (middle and bottom),
which correspond to the leading orders of the energy splitting due to $\mc{H}_{xy}$ (see text).
For $s=1/2$, the dominant off-diagonal processes split completely the 9-fold degeneracy of the Ising point. 
For higher spins, the diagonal processes dominate and give rise to a splitting of $2 v(s,\alpha)$ between the two 
``uud'' families of Eq.~(\ref{CubocIsingRed.Eq}) as predicted from perturbation theory (see text). 
Filled squares denote the eigenvalues of the effective QDM of the corresponding leading term for each $s$.}
\label{Cuboc_XXZ.Fig}
\end{figure}
%%%%%%%%%%%%%%%%%%%%%%%%%%%%%

The 9 ``uud'' states are highlighted in Fig.~\ref{Cuboc_XXZ.Fig} which shows our symmetry-resolved ED results
for $s=1/2$, $3/2$, and $5/2$ (at their $M=M_s/3$ sector) as a function of the anisotropy parameter $\alpha$. 
The energies are given in units of $J_z\alpha^2$ and $J_z\alpha^4$ for $s=1/2$ and $s=3/2,5/2$ respectively, 
which are the leading orders of the energy splitting due to $\mc{H}_{xy}$ (see below). 
Figure~\ref{Cuboc_XXZ.Fig} shows that the Heisenberg states which are adiabatically connected to the lowest Ising 
manifold are not the lowest excitations for $s=1/2$ while this is clearly the case for higher spins and, as we show below,
for the $s=1/2$ icosidodecahedron as well.

We shall try now to understand some of the features of Fig.~\ref{Cuboc_XXZ.Fig} in more detail by considering the lowest 
order effect of $\mc{H}_{xy}$ in splitting the Ising nine-fold degenerate manifold, as a function of spin and $\alpha$. 
We follow the general guidelines and considerations of the Appendix~\ref{DPT.App}.
We distinguish between diagonal and off-diagonal processes depending on whether the initial state is finally
recovered or not. The former come from the smallest closed paths on the molecule (beyond triangles),
which in the present case are the square loops (see left panel of Fig.~\ref{CubocIcosi.Fig})
and the corresponding amplitudes scale with the fourth power of $\alpha$ for all $s$.
Now, there are only three possible configurations on a square which respect the ``uud'' constraint and these are depicted 
in Fig.~\ref{DiagSquare.Fig}(a), (b) and (c). Each one carries a certain diagonal energy, say $\epsilon_a$, $\epsilon_b$ 
and $\epsilon_c$. We have calculated these energies as a function of $s$ using Eq.~(\ref{DPT.Eq}) and by enumerating all
relevant processes. The results are
\begin{eqnarray}
\epsilon_a(s)&=&0~,\nonumber\\
\epsilon_b(s)&=& -\frac{s^3}{2(4 s-1)^2} J_z\alpha^4~,\nonumber\\
\epsilon_c(s)&=& -2\frac{s^4 \delta_{s,1/2}}{(4 s-1)^2(2 s-1)} J_z \alpha^4~.
\end{eqnarray}
The Kronecker symbol $\delta_{s,1/2}$ appears in $\epsilon_c(s)$ because some of the diagonal processes
relevant for $s>1/2$ are not present for $s=1/2$ since they involve intermediate states which do not belong to the
Ising manifold.\footnote{This particular point for $\epsilon_c(s)$
has been overlooked in Ref.~\onlinecite{Bergman2} in the context of the checkerboard lattice.}
Now, to any given Ising configuration $i$ there corresponds an associated potential energy equal to
$E^i=n_a^{i}\epsilon_a + n_b^{i}\epsilon_b + n_c^{i}\epsilon_c$,
where $n_a^i$, $n_b^i$, $n_c^i$ are the number of squares in the states $a$, $b$ and $c$,
respectively in $i$. On the other hand, we must satisfy two global conditions, one for the total number of
squares $N_s=6=n_a^i+n_b^i+n_c^i$, and another for the total number of down spins $N_d=4=n_b^i/2 + n_c^i$. 
This leaves us with one independent, non-global variable, say $n_c^i$, in terms of which one can express $E^i$. 
Omitting a global energy term $8\epsilon_b-2\epsilon_a$, we find $E^i=v(s,\alpha) n_c^i$ with
\be\label{CubocV.Eq}
v(s,\alpha)=\epsilon_a-2\epsilon_b+\epsilon_c=
\frac{s^3 (2 s \delta_{s,1/2}-1)}{(4 s-1)^2(2 s-1)} J_z \alpha^4~,
\ee 
which is positive for $s=1/2$ and negative otherwise. The corresponding (lowest order) effective diagonal Hamiltonian reads 
\begin{eqnarray}\label{CubocQDM_V.Eq}
\mc{V}_{\text{eff}}^{(4)} = v(s,\alpha) \sum
\Big|
\parbox{0.2in}{\epsfig{file=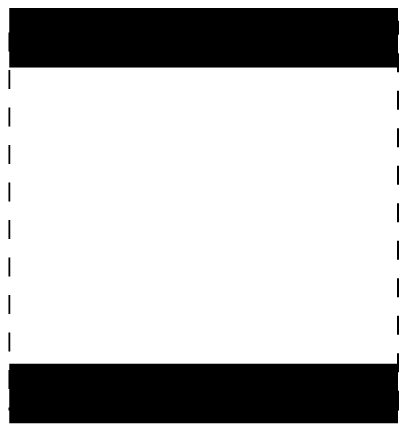,width=0.2in,clip=}}
\Big\rangle
\Big\langle
\parbox{0.2in}{\epsfig{file=CubeQDM_ket,width=0.2in,clip=}}
\Big|~,
\end{eqnarray}
where the sum runs over all six square plaquettes (with both orientations) of the cluster.
We may easily check that among all nine possible dimer coverings of the cube, three of them have $n_c=4$ 
and thus $\mc{V}_{\text{eff}}^{(4)}=4 v(s,\alpha)$, while the remaining six have $n_c=2$ 
and thus $\mc{V}_{\text{eff}}^{(4)}=2 v(s,\alpha)$. 
These correspond to the two families of ``uud'' states mentioned above, see Eq.~(\ref{CubocIsingRed.Eq}). 
Hence, the eigenvalues of $\mc{V}_{\text{eff}}^{(4)}$ form a pair of a 6-fold and a 3-fold degenerate levels 
with an energy splitting of $2 v(s,\alpha)$ between them. In particular, this splitting amounts to 
$\frac{27}{200} J_z\alpha^4$ for $s=3/2$ and $\frac{125}{1296} J_z\alpha^4$ for $s=5/2$.

%%%%%%%%%%%%%%%%%%%%%%%%%%%%%
\begin{figure}
\centering
\includegraphics*[width=0.6\linewidth]{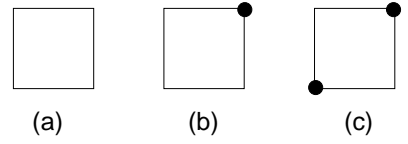}
\caption{The three possible local (on the square loops) views of the ``uud'' configurations of the
cuboctahedron. Filled circles denote spins with $m=-s$, all remaining vertices have $m=s$.}
\label{DiagSquare.Fig}
\end{figure}
%%%%%%%%%%%%%%%%%%%%%%%%%%%%%

We now consider off-diagonal processes. To lowest order in $\alpha$, these are confined to the maximally
flippable even-length loops of the cluster. These loops are the alternating spin up-down configurations
already shown in Fig.~\ref{CubocOffDiagLoop.Fig}. The corresponding flipping amplitude $t$ scales as $\alpha^{4 s}$. 
Their explicit values for several $s$ are provided in Table~\ref{Amplitudes.Table} (with $L=4$) of Appendix~\ref{DPT.App}.
Thus the leading kinetic effect is described in the dimer representation by the term
\begin{eqnarray}\label{CubocQDM_T.Eq}
\mc{T}_{\text{eff}}^{(4 s)} &=& t(s,\alpha) \sum\left(
\Big|
\parbox{0.2in}{\epsfig{file=CubeQDM_ket,width=0.2in,clip=}}
\Big\rangle
\Big\langle
\parbox{0.2in}{\epsfig{file=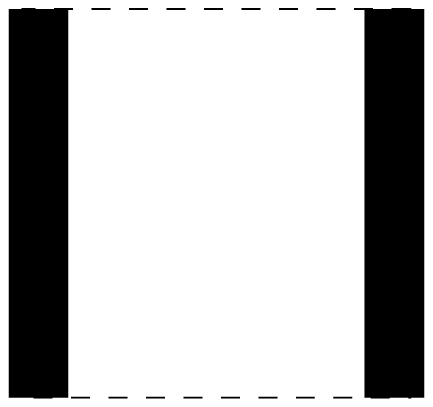,width=0.2in,clip=}}
\Big|+ \text{h.c.} \right)~,
\end{eqnarray}
where $t(s,\alpha)$ can be calculated explicitly using Eq.~(\ref{DPT.Eq}) and enumerating all different processes. 
Several representative values are provided in Table~\ref{Amplitudes.Table}. The eigenvalues of 
$\mc{T}_{\text{eff}}^{(4 s)}$ in units of $t$ are:
$-2\sqrt{2},-\sqrt{2}, -\sqrt{2}, 0, 0, 0, \sqrt{2}, \sqrt{2}, 2\sqrt{2}$. These correspond to a complete splitting 
of the different IR's of each of the two families of Eq.~(\ref{CubocIsingRed.Eq}). 

According to the above, our effective quantum dimer model should generally include both kinetic and potential terms. 
To leading order, this model reads
\be\label{CubocQDM.Eq}
\mc{H}_{\text{eff}}=\mc{V}_{\text{eff}}^{(4)} + \mc{T}_{\text{eff}}^{(4 s)}~.
\ee 
For $s=1/2$ we have $t=-J_z\alpha^2$, $v=J_z\alpha^4/8$ and the dynamics is mainly governed by kinetic processes which 
split completely (cf. Fig.~\ref{Cuboc_XXZ.Fig}(a)) the nine ``uud'' states. 
For $s=1$ we have $t=-J_z\alpha^4$, $v=-J_z\alpha^4/9$ and thus both diagonal and off-diagonal
processes are equally important. For $s>1$ the diagonal processes dominate and give rise to a splitting of $2 v(s,\alpha)$ 
between $\mathbf{P}^{(6)}_{\text{uud}}$ and $\mathbf{P}^{(3)}_{\text{uud}}$ of Eq.~(\ref{CubocIsingRed.Eq}). 
In particular, since $v<0$, the states of $\mathbf{P}^{(3)}_{\text{uud}}$ will be favored because they have a larger number 
(four) of the plaquettes of Fig.~\ref{CubocOffDiagLoop.Fig}(c). 
All these features are nicely demonstrated in Fig.~\ref{Cuboc_XXZ.Fig} where we compare our ED results for the XXZ model 
at small $\alpha$ with the leading order eigenvalues of Eq.~(\ref{CubocQDM.Eq}) which are shown as (red) filled squares.

\subsubsection{Icosidodecahedron}\label{Icosi_Plateau.Subsubsec}
\begin{figure}
\centering
\includegraphics*[width=0.6\linewidth]{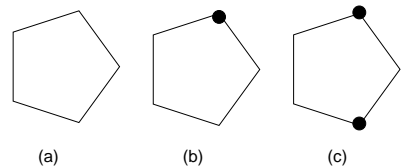}
\caption{The three possible local (on the pentagonal loops) views of the ``uud'' configurations on the
icosidodecahedron. The filled circles denote spins with $m=-s$, all remaining vertices have $m=s$.}
\label{DiagPentagon.Fig}
\end{figure}

We turn now to the corresponding plateau phase of the Heisenberg icosidodecahedron and follow a similar analysis as above. 
Here the lowest Ising manifold, henceforth $\mathbf{R}_\textrm{uud}$, consists of 36 ``uud'' states which are in one-to-one 
correspondence with the 36 dimer coverings of the dodecahedron. 
We find that this manifold decomposes into two invariant (under $\mathsf{I}_h$) families 
$\mathbf{R}^{(30)}_{\text{uud}}$ and $\mathbf{R}^{(6)}_{\text{uud}}$ of 30 and 6 states respectively as
\be\label{IcosiIsingRed.Eq}
\mathbf{R}_\textrm{uud} = \mathbf{R}^{(30)}_{\text{uud}}\oplus\mathbf{R}^{(6)}_{\text{uud}}~,  
\ee
where 
\begin{eqnarray}
\mathbf{R}^{(30)}_{\text{uud}} &=& \text{Ag}\oplus\text{Au}\oplus\text{Fg}\oplus\text{Fu}\oplus 2\text{Hg}\oplus 2\text{Hu}~,
\nonumber\\ 
\mathbf{R}^{(6)}_{\text{uud}} &=& \text{Ag}\oplus\text{Hg}~.
\end{eqnarray}
Each of these families contains states with a fixed number $n_c$ (8 and 10 respectively) of pentagonal plaquettes 
of the type of Fig.~\ref{DiagPentagon.Fig}(c) (this number remains invariant under the symmetry operations 
$\mathsf{I}_h$ of the cluster). We should note in particular that each of the 6 states of $\mathbf{R}^{(6)}_{\text{uud}}$ 
contain two pentagons with all spins pointing up (i.e. have $n_a=2$, cf.Fig.~\ref{DiagPentagon.Fig}(a)).

\begin{figure}
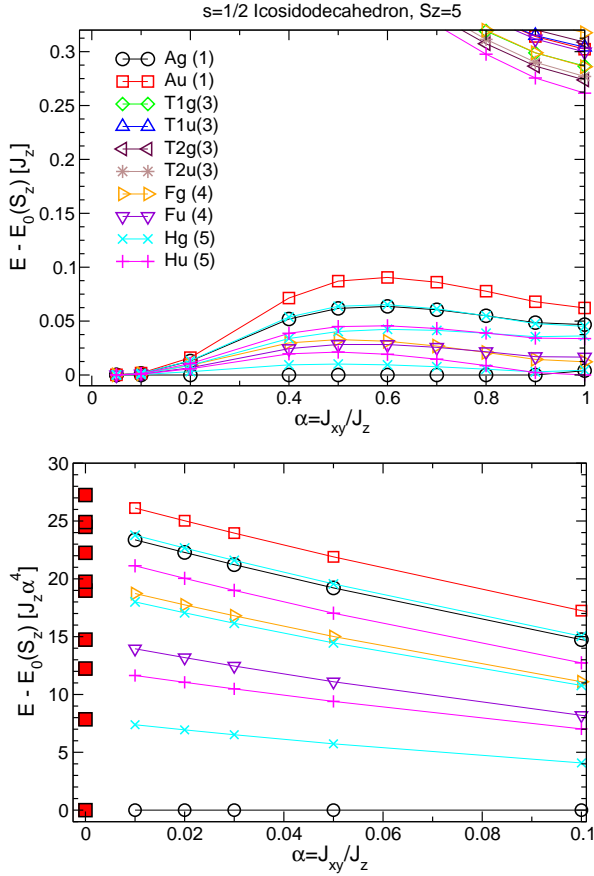

\centering
\includegraphics*[width=0.9\linewidth]{IcosiOneHalf_XXZ_Full}\\
\vspace{2mm}
\includegraphics*[width=0.9\linewidth]{IcosiOneHalf_XXZ_vs_Dimer}
\caption{(Color online) (a) Lowest eigenvalues of the $s=1/2$ XXZ model on the icosidodecahedron
in the $S_z=5$ sector. Interpolation between the $\alpha=0$ Ising and the $\alpha=1$ Heisenberg point.
The large energy separation between the lowest 36 states and higher excitations is clearly evident.
(b) Convergence of the lowest 36 eigenvalues (in units of $J_z\alpha^4$ which is the leading order)
toward the eigenvalues (filled squares) of the effective dimer Hamiltonian 
$\mc{T}_{\text{eff}}^{(4)}$ of Eq.~(\ref{Icosi_Heff.Eq}) as described in the text.}
\label{Icosi_XXZ.Fig}
\end{figure}

A simple inspection of Fig.~\ref{Icosi_XXZ.Fig}(a), which shows the $s=1/2$ lowest energy spectrum of the XXZ model, 
reveals that the lowest 36 Heisenberg states trace back to the ground state ``uud'' manifold of the Ising point. 
A striking difference to the $s=1/2$ cuboctahedron case studied above is that as these lowest Ising states ``evolve'' 
toward their low-lying Heisenberg counterparts, they remain always well separated from the higher energy states 
in the $Sz=5$ subspace. 
We now give a complementary picture, which is valid at least for small $\alpha$, by considering the lowest order 
processes driven by $\mc{H}_{xy}$ and by deriving the corresponding effective QDM on the dodecahedron. 
We begin with the lowest order off-diagonal processes. 
As above, these stem from maximally flippable loop configurations of the smallest possible even length $L$. 
Such a loop configuration that respects the local ``uud'' constraint is the octagonal loop with alternating up-down spins 
depicted in Fig.~\ref{IcosiOffDiagLoop.Fig}(a) which in turn maps to the flippable plaquette of the dodecahedron shown
in Fig.~\ref{IcosiOffDiagLoop.Fig}(b).
For spin $s=1/2$ then, the lowest off-diagonal term in $\mc{H}_{\text{eff}}$ is of fourth order in
$\mc{H}_{xy}$. Each ``up-down'' loop of the type shown in Fig.~\ref{IcosiOffDiagLoop.Fig}(a) is amenable to a
kinetic process of the form $t~s_1^-s_2^+s_3^-\ldots s_8^+$. 
From our calculations, shown in Table~\ref{Amplitudes.Table}, we obtain $t=-2.5 J_z\alpha^4$.
In fifth order, we find two types of kinetic processes. The first is similar to the above but
now involves loops of length $10$ such as the equators of the molecule.
The second type is less obvious, and invokes again the octagonal loops of Fig.~\ref{IcosiOffDiagLoop.Fig}(a)
and any one of the neighboring spin sites. 
Since these loops map to exactly the same flippable dimer plaquette of Fig.~\ref{IcosiOffDiagLoop.Fig}(b)
their effect is to merely renormalize the fourth-order amplitude $t$.
In fact, these terms result in an overall decrease of $|t|$ since they carry an extra negative sign.

%%%%%%%%%%%%%%%%%%%%%%%%%%%%%
%%%%%%%%%%%%%%%%%%%%%%%%%%%%%

On the other hand, the lowest order diagonal processes must be confined to the smallest closed path
which in the present case are the pentagons of the molecule (see right panel of
Fig.~\ref{CubocIcosi.Fig}). The three possible types of configurations that respect the ``uud''
constraint around a pentagon are depicted in Fig.~\ref{DiagPentagon.Fig}
and are designated by $a$, $b$ and $c$. Each one carries a certain diagonal energy,
say $\epsilon_a$, $\epsilon_b$ and $\epsilon_c$. These are calculated by enumerating all relevant processes
and using Eq.~(\ref{DPT.Eq}) of Appendix~\ref{DPT.App}. They are explicitly given by
\begin{eqnarray}
\epsilon_a &=& 0~,\nonumber\\
\epsilon_b &=& \frac{s^3}{8 (4 s-1)^2} J_z\alpha^5~,\nonumber\\
\epsilon_c &=& \frac{6 s^4}{(4 s-1)(8 s-3)^2} J_z\alpha^5~.
\end{eqnarray}
To any given Ising configuration $i$ there corresponds an associated energy equal to
$E^i=n_a^{i}\epsilon_a + n_b^{i}\epsilon_b + n_c^{i}\epsilon_c$,
where $n_a^i$, $n_b^i$, $n_c^i$ are the number of pentagons in the states $a$, $b$ and $c$
respectively in $i$. On the other hand, we must again satisfy two global conditions, one for the total
number of pentagons $N_p=12=n_a^i+n_b^i+n_c^i$, and one for the total number of down spins
$N_d=10=n_b^i/2 + n_c^i$. This leaves us with one independent, non-global variable, which we choose to be
$n_c^i$. Omitting a global energy $20\epsilon_b-8\epsilon_a$, we find $E^i=v(s,\alpha) n_c^i$, with
\be\label{Icosi_V.Eq}
v(s,\alpha)=\epsilon_a-2\epsilon_b+\epsilon_c = \frac{s^3 (32 s^2+24 s-9)}{4 (4 s-1)^2(8 s-3)^2} J_z \alpha^5~,
\ee
which is positive for all $s$. This means that $v(s,\alpha)$ favors configurations with the minimum number
of the pentagonal states of Fig.~\ref{DiagPentagon.Fig}(c). 
Since all 30 states of $\mathbf{R}^{(30)}_{\text{uud}}$ have $n_c=8$ while the 6 states of 
$\mathbf{R}^{(6)}_{\text{uud}}$ have $n_c=10$, the former family will be lower in energy by a splitting 
of $2 v(s,\alpha)$. Furthermore, it is clear that diagonal processes do not give rise to a 
splitting within the two families. 

%%%%%%%%%%%%%%%%%%%%%%%%%%%%%
%%%%%%%%%%%%%%%%%%%%%%%%%%%%%
%%%%%%%%%%%%%%%%%%%%%%%%%%%%%
\begin{table}
\caption{Eigenvalues (in units of $t$) of $\mc{H}_{\text{eff}}$ given in Eq.~(\ref{Icosi_Heff.Eq}),
together with their multiplicities.}\label{DimerEigenvalues.Table}
\begin{ruledtabular}
\begin{tabular}{ll||ll}
Energy$[t]$ & deg. & Energy$[t]$ & deg.\\
\hline
-4        & 1 & -0.694593 & 5\\
-3.06418  & 5 & 1         & 4\\
-2.89898  & 1 & 2         & 5\\
-2        & 5 & 3.75877   & 5\\
-1        & 4 & 6.89898   & 1\\
\end{tabular}
\end{ruledtabular}
\end{table}
%%%%%%%%%%%%%%%%%%%%%%%%%%%%%

\textit{$s=1/2$ case}.---
Given all the above, the effective QDM for the plateau phase of the $s=1/2$ Heisenberg
icosidodecahedron is, at lowest order, dominated by kinetic, off-diagonal processes of the form
\be\label{Icosi_Heff.Eq}
\mc{H}_{\text{eff}} \simeq \mc{T}^{(4)}_{\text{eff}}  = 
t\sum\Big|
\parbox{0.36in}{\epsfig{file=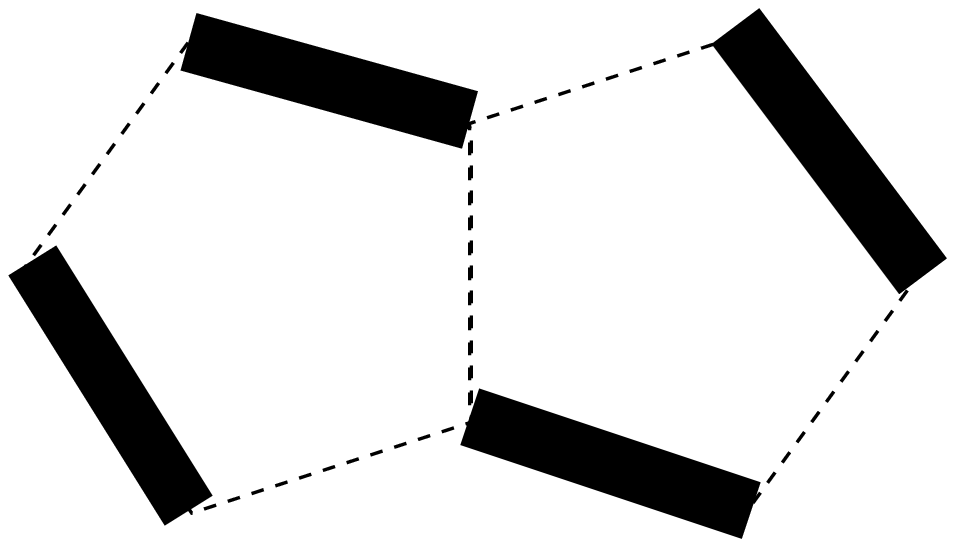,width=0.36in,clip=}}
\Big\rangle
\Big\langle
\parbox{0.36in}{\epsfig{file=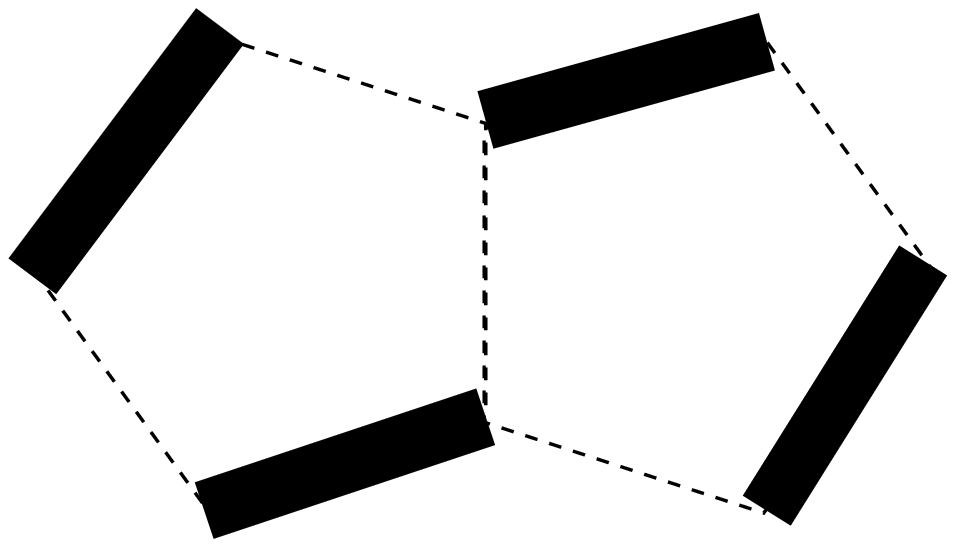,width=0.36in,clip=}}
\Big|+ \text{h.c.}~,
\ee
where the sum runs over all octagonal plaquettes and $t=-2.5 J_z\alpha^4$. 
The corresponding $36 \times 36$ Hamiltonian matrix can be constructed and solved numerically for its eigenvalues. 
These are provided in Table~\ref{DimerEigenvalues.Table} in units of $t$.
They are also shown as filled squares in Fig.~\ref{Icosi_XXZ.Fig}(b).
In the same figure, we show the lowest 36 eigenvalues (divided by $J_z$) of $\mc{H}'$
in units of $\alpha^4$. The clear convergence for small $\alpha$ toward the eigenvalues of
$\mc{H}_{\text{eff}}$ confirms the validity of our perturbative calculations. Moreover, the fact that
the convergence is linear confirms that the next processes contributing to $\mc{H}_{\text{eff}}$ come
in fifth order. In particular, the clear decrease of the bandwidth with $\alpha$ is in agreement with our
previous assertion that the fourth order amplitude $t$ of Eq.~(\ref{Icosi_Heff.Eq}) gets renormalized
from the fifth order octagonal kinetic processes mentioned above.
The latter seem to dominate over the corresponding fifth order off-diagonal decagonal loop processes
and the fifth order diagonal ones. Looking at Fig.~\ref{Icosi_XXZ.Fig}(a) one notes that this may be even
more general: To all orders in $\mc{H}_{xy}$, there seems to be a mere renormalization of the bandwidth without drastically
altering the relative amplitudes of the eigenvalues of $\mc{T}_{\text{eff}}^{(4)}$.
This suggests a dominance of the most local (octagonal) kinetic processes renormalized from all orders in $\mc{H}_{xy}$.

\textit{$s>1/2$ and relevance to Mo$_{72}$Fe$_{30}$}.---
As mentioned above, diagonal processes first appear in fifth order irrespective of $s$. On the other hand,
the lowest off-diagonal process on a loop of $L$ (even) sites appears in order $L s$.
For instance, the octagonal loops discussed above give processes at order $8 s$ (and higher), whereas
decagonal (e.g. the equatorial) loops contribute in order $10 s$ (and higher). Hence for $s=1$ diagonal and
off-diagonal processes are equally important, while for $s>1$ the physics will be completely dominated by 
diagonal processes of fifth order in $\alpha$:
\be\label{Icosi_Heff_diag.Eq}
\mc{H}_{\text{eff}} \simeq \mc{V}^{(5)}_{\text{eff}} = 
v(s,\alpha)\sum\Big|
\parbox{0.25in}{\epsfig{file=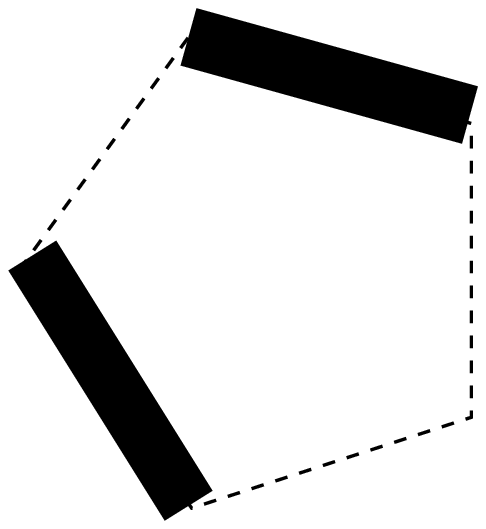,width=0.25in,clip=}}
\Big\rangle
\Big\langle
\parbox{0.25in}{\epsfig{file=DodecaDiag_ket,width=0.25in,clip=}}
\Big|~,
\ee
where the sum runs over all pentagonal plaquettes (with all possible orientations) and $v(s,\alpha)$ 
is given by Eq.~(\ref{Icosi_V.Eq}).
As explained above these processes result in a diagonal splitting of $2 v(s,\alpha)$ between  
$\mathbf{R}_{\text{uud}}^{(30)}$ and $\mathbf{R}_{\text{uud}}^{(6)}$ with the former family being lowest in energy. 
For $s=5/2$ in particular, which is relevant for the plateau phase of Mo$_{72}$Fe$_{30}$~\cite{Fe30_Plateau}, 
this splitting amount to the quite small value of $2 v\simeq 0.0838 J_z\alpha^5$.
Since, at the same time, the excitations out of the Ising manifold are expected to remain gapped for the finite value 
of $s=5/2$, we speculate that the plateau phase of Mo$_{72}$Fe$_{30}$ must show a characteristic low-temperature
thermodynamic signal which is qualitatively similar to that of Fig.~\ref{EntropySpecHeat.Fig}
(up to fine details related to the 30 to 6 diagonal splitting). Here, in particular, the entropy content
of the renormalized ``uud'' manifold amounts to approximately $7\%$ of the full magnetic entropy. 
This calls for low-temperature specific heat measurements on Mo$_{72}$Fe$_{30}$ at the plateau phase as explained 
in Sec.~\ref{Thermo.Subsec}. More generally, it is exciting that the notion of a Quantum Dimer Model finds a realization
in the low-energy $M=M_s/3$ plateau physics of finite-size magnetic clusters like Mo$_{72}$Fe$_{30}$ or Mo$_{72}$V$_{30}$.

\section{Heisenberg spectra for $s>1/2$}\label{Coplanarity.Sec}
\begin{figure}[!t]
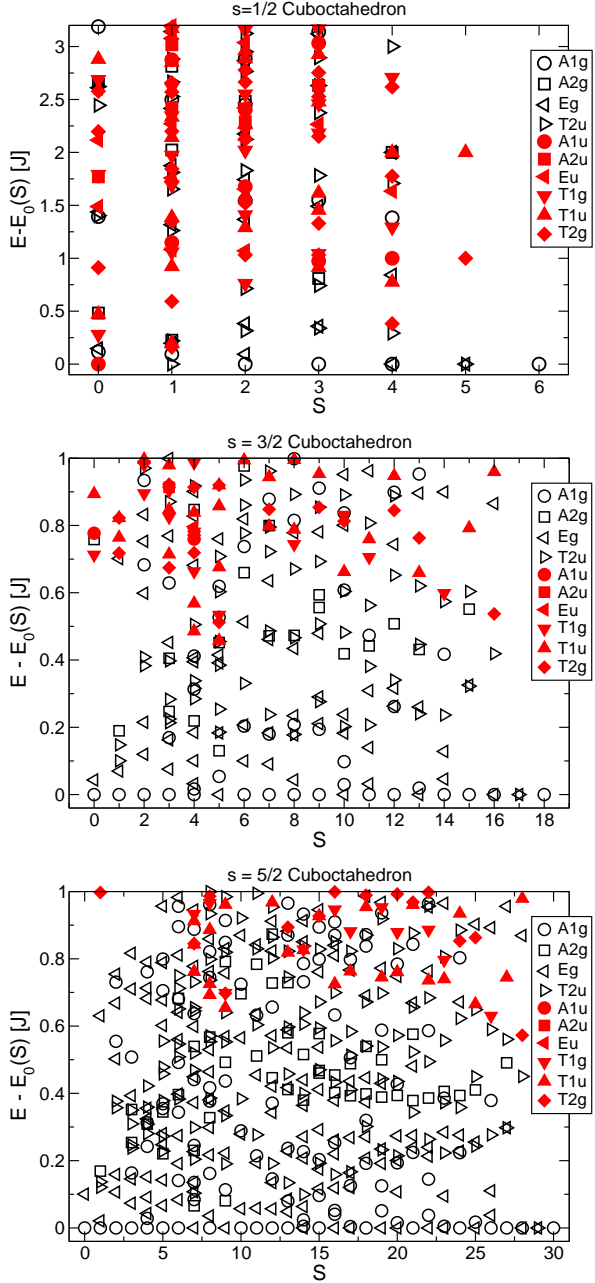

\centering
\includegraphics*[width=0.9\linewidth]{CubocOneHalf_HeisStot}\\
\vspace{2mm}
\includegraphics*[width=0.9\linewidth]{CubocThreeHalf_HeisStot}\\
\vspace{2mm}
\includegraphics*[width=0.9\linewidth]{CubocFiveHalf_HeisStot}
\caption{(Color online) Low-energy spectra of the AFM Heisenberg cuboctahedron for $s=1/2$ (top),
$3/2$ (middle) and $5/2$ (bottom) in terms of the total spin $S$.
All energies are measured from their corresponding $E_0(S)$.}
\label{Cuboc_Heis.Fig}
\end{figure}

The major focus of this section is on Heisenberg spectra with $s>1/2$. Our interest in this regard is mainly motivated 
by the INS experiments reported by Garlea \textit{et al.}~\cite{Garlea_INS} on Mo$_{72}$Fe$_{30}$. The main finding 
of these experiments is a very broad response which manifests in a wide range of fields. Previous theories which are based 
either on the excitations of the rotational band model~\cite{Garlea_INS,SchnackLubanModler} or on spin wave 
calculations~\cite{Cepas_LSW,Waldmann_LSW}, could not account for the observed behavior since they predict only a small 
number of discrete excitation lines at low temperatures. Although the diagonalization of the $s>1/2$ icosidodecahedron 
is not feasible (at low magnetizations) with current computational power, an immediate interpretation of this behavior 
can be deduced from a study of the cuboctahedron. 
Exact diagonalization spectra of this cluster for $s=3/2$ and $s=5/2$ are shown 
in the two lower panels of Fig.~\ref{Cuboc_Heis.Fig}. For comparison we have also included the $s=1/2$ spectrum 
(this was shown before in the middle panel of Fig.~\ref{12mer_Cuboc_Icosi_Heis.Fig} in terms of $S_z$ instead of the total spin $S$). 
Here, in contrast to unfrustrated clusters 
(cf. upper panel of Fig.~\ref{12mer_Cuboc_Icosi_Heis.Fig}), there does not exist 
any well isolated and thus clearly identified low-energy tower of states or rotational band. Instead, a ``bulk'' 
of very dense excitations are present in the full magnetization range. This is a generic feature of highly 
frustrated systems which manifests irrespective of $s$ and thus must be also present in the $s=5/2$ Mo$_{72}$Fe$_{30}$ 
cluster (similarly to its $s=1/2$ analogue of Fig.~\ref{12mer_Cuboc_Icosi_Heis.Fig}).

The origin of these dense excitations for $s>1/2$ can be readily suggested by the following striking observation 
in Fig.~\ref{Cuboc_Heis.Fig}: The spectra consist, up to a relatively large energy cut-off, 
entirely of the representations A1g, A2g, Eg, and T2u.  
The main message in the following is that this peculiar spatial symmetry pattern as well as the 
combined spatial$+$spin pattern (i.e., the appearance of specific sets of spatial IR's in each $S$ sector) 
are characteristic fingerprints of the 3-sublattice Heisenberg classical GS's. For instance, as we explain below, 
these four representations are exactly the ones that appear in the symmetry decomposition of 
the coplanar classical GS's (cf. Eq.~(\ref{CubocRed.Eq}) below). The dense excitation features of the lower two panels 
of Fig.~\ref{Cuboc_Heis.Fig} can be thereby accounted for by the large spatial degeneracy of these configurations, 
a fact whose importance does not seem to have been recognized in the past. 
In principle, each of these states gives rise to a distinct ``tower of states'' or ``rotational band'' 
and they all appear together at low energies albeit split by quantum fluctuations. So the large discrete degeneracy has 
a direct impact on the low-energy spectrum, and we believe this large number of levels is at the very heart of the broad 
INS response reported in Ref.~\onlinecite{Garlea_INS}. By contrast, the absence of a clear symmetry pattern in the 
$s=1/2$ spectra (cf. upper panel of Fig.~\ref{Cuboc_Heis.Fig}) shows that the associated low-lying excitations 
are of different origin. 

%%%%%%%%%%%%%%%%%%%%%%%%%%%%%
%%%%%%%%%%%%%%%%%%%%%%%%%%%%%

Before analyzing further our numerical results it is useful to recall (cf. Subsec.~\ref{kagome.Subsec} below) 
what is known about the classical GS's of the infinite kagom\'e lattice in zero and finite field and discuss 
what carries over in the present clusters. In particular, we give the explicit spatial degeneracy of the coplanar GS's 
and a short summary of their symmetry properties. The latter have been derived independently by employing a group 
theoretical analysis, the details of which have been relegated to Appendix~\ref{Tower.App}. 
Our semiclassical interpretation for the origin of the dense excitations of the above $s>1/2$ 
spectra will be further corroborated by a closer comparison of the symmetry pattern of the spectra 
with the combined spatial$+$spin symmetry of the coplanar GS's also derived in Appendix~\ref{Tower.App}. 
In Subsec.~\ref{XY.Subsec} we discuss the simpler case of the $s>1/2$ XY model which exemplifies 
very evidently the main idea of this section, i.e., the simultaneous 
presence of several lowest towers of states due to the discrete classical degeneracy. 
%%%%%%%%%%%%%%%%%%%%%%%%%%%%%
\begin{figure}%[!t]
\centering
\includegraphics*[width=0.9\linewidth]{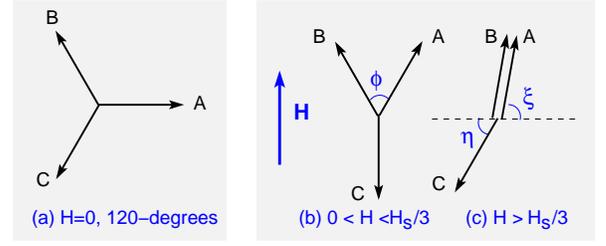}
\caption{Classical GS's of the Heisenberg model on the kagom\'e AFM which are selected by quantum or thermal
fluctuations. (a) In zero-field, these are the 120$^\circ$ states. In finite fields ((b) and (c)), the three spins 
lie on the plane of the field. (b) For $H < H_s/3$, we have a one-parameter ($\phi < 120^\circ$) family of states 
with one of the spins (C) pointing antiparallel to the field. (c) For $H\geq H_s/3$, we have a two-parameter 
($\xi$ and $\eta$) family of states with two spins (A and B) being collinear.}\label{CoplanarStates.Fig}
\end{figure}
%%%%%%%%%%%%%%%%%%%%%%%%%%%%%

\subsection{Classical GS's and large spatial degeneracy}\label{kagome.Subsec}
Let us first consider the ground state configurations of the classical Heisenberg model in the infinite 
kagom\'e system and the present clusters. The corner-sharing triangles structure makes the discussion rather simple. 
The classical Hamiltonian can be rewritten in the suggestive form (in units of $g\mu_B=1$)
\be\label{HeisClassical.Eq}
\mathcal{H}^\mathrm{HB}_\mathrm{classical} = \frac{J}{2}\ \sum_{\Delta} \left(\mathbf{S}_\Delta - \vec{H}/2J\right)^2,
\ee
where $\mathbf{S}_\Delta$ denotes the total spin on a triangle $\Delta$. In this form it is straightforward to
see that all configurations with $\mathbf{S}_\Delta = \vec{H}/2J$ on each triangle are GS's. 
It is useful to examine the zero-field case first.

\textit{Zero-field case}.---
Here, the classical constraint $\mathbf{S}_\Delta = 0$ amounts to a simple 120$^\circ$ configuration of the three spins, 
which is depicted in Fig.~\ref{CoplanarStates.Fig}(a). 
An important point here and in the following is to determine {\em how many} such GS's exist. 
For the kagom\'e lattice it is well known that the 
ground state manifold consists of both coplanar and non-coplanar configurations in zero field. The coplanar GS's 
are extensively degenerate as can be shown by a mapping onto vertex three-colorings of the kagom\'e lattice or 
equivalently onto bond three-colorings of the Honeycomb lattice~\cite{Potts,BaxterThreeColorings}. On the other hand 
the non-coplanar GS's can be generated from the coplanar ones by the following recipe: 
Identify a loop of alternating spin orientations (two out of three directions), which is either closed
or extends to infinity. All sites neighboring the loop share the common third spin direction. It is then possible
to collectively rotate the spins on the loop freely around the third direction at zero energy cost.
Such a new state is clearly non-coplanar, but still a ground state. 

%%%%%%%%%%%%%%%%%%%%%%%%%%%%%
\begin{figure}%[!t]
\centering
\includegraphics*[width=0.9\linewidth]{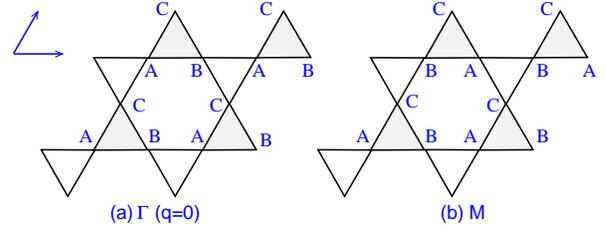}
\caption{Two of the 24 possible vertex three-colorings on the cuboctahedron
which is projected on a plane here (periodic boundary conditions along the two arrows are implied).
(a) One of the 6 colorings of the $\Gamma$ (or $q=0$) family, and (b) one of the 18 colorings of the $M$ family.}
\label{CubocOnThePlane.Fig}
\end{figure}
%%%%%%%%%%%%%%%%%%%%%%%%%%%%%

Let us now discuss what carries over of this large classical degeneracy on the two present molecules.
It is straightforward to enumerate all vertex-three colorings for both the cuboctahedron and the icosidodecahedron
and the respective numbers are 24 and 60, or 4 and 10 if one discards global recolorings. Two such states of the 
cuboctahedron are illustrated in Fig.~\ref{CubocOnThePlane.Fig}, while a typical one for the icosidodecahedron 
can be found in Fig.~2 of Ref.~\onlinecite{AxenovichLuban}. We stress here that these 4 and 10 states are unrelated 
by the global $\mathsf{O}(3)$ symmetry, and therefore form genuinely different GS's and give rise to distinct 
``tower of states'' at low-energies. The spatial symmetry properties of these states have been derived 
in Appendix~\ref{Tower.App} and can be summarized as follows. The 24 vertex three-colorings of the cuboctahedron 
form two invariant (under the operations of $\mathsf{O}_h$) families $\mathbf{P}^\Gamma_{\text{ABC}}$ 
and $\mathbf{P}^M_{\text{ABC}}$ which consist of 6 and 18 states respectively, and they decompose into IR's 
of $\mathsf{O}_h$ as
\begin{eqnarray}\label{CubocRed.Eq}
\mathbf{P}_{\text{ABC}}^{\Gamma} &=& \text{A1g} \oplus \text{A2g} \oplus 2\text{Eg}~,\nonumber\\
\mathbf{P}_{\text{ABC}}^{M} &=& 3(\text{A1g} \oplus \text{Eg} \oplus \text{T2u})~.
\end{eqnarray}
As mentioned above, these are exactly the IR's that appear (with open symbols) in the spectra of the lower two panels of 
Fig.~\ref{Cuboc_Heis.Fig}. 
On the other hand, the 60 vertex three-colorings of the icosidodecahedron form two invariant (under the operations of 
$\mathsf{I}_h$) families which are equivalent to each other (cf. Appendix~\ref{Tower.App}).
Here, we shall treat these families collectively as a 
single one called $\mathbf{R}_{\text{ABC}}$. This decomposes into IR's of $\mathsf{I}_h$ 
as~\footnote{We should note here that larger clusters have a larger number of coplanar GS's 
and thus decompose into an accordingly larger number of spatial IR's (eventually containing all possible spatial IR's 
for large enough sizes). This can be already seen for the 
icosidodecahedron whose classical GS's decompose into 8 out the 10 different IR's of $\mathsf{I}_h$. 
The cuboctahedron on the other hand, has a small number of coplanar GS's and this allows to recognize 
their traces in the low-lying excitation spectra.}
\begin{eqnarray}\label{IcosiRed.Eq}
\mathbf{R}_{\text{ABC}} = 2 (\text{Ag} \oplus \text{Au} \oplus \text{Fg}
\oplus \text{Fu} \oplus 2\text{Hg} \oplus 2\text{Hu})~.
\end{eqnarray}  

As to the non-coplanar GS's (in zero-field), it turns out that the icosidodecahedron has none, 
since all the alternating loops described above have maximal length (20), and therefore the rotation of the 
spins on the loop just changes the global spin plane, thus preserving the coplanarity. 
The cuboctahedron however has non-coplanar GS's, since the loops can have length shorter than eight, 
in agreement with previous studies~\cite{AxenovichLuban,SchmidtLuban}.

When switching on quantum fluctuations on the kagom\'e lattice it is known that the spin waves at harmonic order select 
the coplanar GS's over the non-coplanar ones (order-by-disorder effect), due to the larger number of soft 
modes of the former~\cite{HarrisKallinBerlinsky}. A complete lifting of the remaining (spatial) degeneracy is taking place 
at the level of anharmonic spin waves, whereby the 
single $\sqrt{3}\times\sqrt{3}$ magnetically ordered state is selected~\cite{Chubukov,ChanHenley}. 
On the other hand for the extreme quantum case of $s=1/2$ a number of numerical works clearly show the absence of 
any magnetic order~\cite{Lecheminant_kagome,Waldtmann}. So based on these conflicting results it is difficult to predict to which 
regime the intermediate values of spin will belong. 
%%%%%%%%%%%%%%%%%%%%%%%%%%%%%
\begin{table}[!t]
\caption{Heisenberg point: Decomposition of semiclassical coplanar states into IR's of
$\mathsf{G}=\mathsf{SU}(2)\times \mathsf{R}$ (where $\mathsf{R}=\mathsf{O}_h$ or $\mathsf{I}_h$) up to $S=6$.
N denotes the number of IR's for each $S$ sector of $\mathsf{SU}(2)$, and is equal to
$(2 S + 1)$ times the number of coplanar states in each family divided by six (i.e., 4 and
10 for the cuboctahedron and the icosidodecahedron respectively).
For the derivation see Appendix~\ref{Tower.App}.\label{PottsHeis.Table}}
\begin{ruledtabular}
\begin{tabular}{l|lr|lr|lr}
    &  Cuboc.       &&Cuboc.&&  Icosi. & \\
$S$ & $\mathbf{P}^\Gamma_{\text{ABC}}(1\times 6)$&N&$\mathbf{P}^M_{\text{ABC}}(3\times 6)$&N&$\mathbf{R}_{\text{ABC}}(10\times 6)$& N\\
\hline\hline
0 & A1g          &1 & A1g,Eg       &3  & Ag,Au,Fg,Fu    &10\\
\hline
1 & A2g,Eg       & & A1g,Eg,      &   & Ag,Au,Fg,Fu,   &  \\
  &              &3  & 2T2u         &9 & 2(Hg,Hu)       &30\\
\hline
2 & A1g,2Eg      & & 3(A1g,Eg),   &   & Ag,Au,Fg,Fu,   &  \\
  &              &5  & 2T2u         &15 & 4(Hg,Hu)       &50\\
\hline
3 & A1g,2A2g,2Eg & & 3(A1g,Eg),   &   & 3(Ag,Au,Fg,Fu),&  \\
  &              &7  & 4T2u         &21 & 4(Hg,Hu)       &70\\
\hline
4 & 2A1g,A2g,3Eg & & 5(A1g,Eg),   &   & 3(Ag,Au,Fg,Fu),&  \\
  &              &9  & 4T2u         &27 & 6(Hg,Hu)       &90\\
\hline
5 & A1g,2A2g,4Eg & & 5(A1g,Eg),   &   & 3(Ag,Au,Fg,Fu),&   \\
  &              &11  & 6T2u         &33 & 8(Hg,Hu)       &110\\
\hline
6 &3A1g,2A2g,4Eg && 7(A1g,Eg),   &   & 5(Ag,Au,Fg,Fu),&   \\
  &              &13  & 6T2u         &39 & 8(Hg,Hu)       &130\\
\end{tabular}
\end{ruledtabular}
\end{table}
%%%%%%%%%%%%%%%%%%%%%%%%%%%%%

We now give a symmetry analysis for the cuboctahedron spectra at low magnetizations which suggests strongly 
that the low-lying excitations can be described in semiclassical terms at the harmonic spin-wave level. 
To this end, we should first emphasize that the non-coplanar GS's do not carry the above spatial symmetry because 
not all triangles share the same spin plane in these configurations and thus spatial operations generally cannot 
relate different triangles. This means that the striking agreement between the spatial symmetry of the above 
low-energy spectra and Eq.~(\ref{CubocRed.Eq}) is not just accidental. Of course there can be other states whose 
spatial decomposition can in principle contain some or all of the IR's of Eq.~(\ref{CubocRed.Eq}) 
(in fact such states will be examined below for finite magnetizations). 
More stringent evidence comes by comparing the full spatial$+$spin symmetry pattern of the exact spectra to that of 
the 120$^\circ$ states. The latter has been derived in Appendix~\ref{Tower.App} and the results are provided 
in Table~\ref{PottsHeis.Table} up to $S=6$. A closer inspection of the $s=3/2$ and $5/2$ spectra 
(cf. lower two panels of Fig.~\ref{Cuboc_Heis.Fig}) shows a remarkable agreement: 
All lowest-energy levels (shown with open (black) symbols) which are below the levels shown with filled (red) symbols 
can be identified in Table~\ref{PottsHeis.Table} with the right combinations of spatial and spin representations 
and multiplicity. It is important to note here that, although these towers are severely split by quantum fluctuations 
---in fact some IR's of Table~\ref{PottsHeis.Table} (e.g. one A1g level at the $S=0$ sector) can be found slightly 
higher than the lowest (red) filled symbols--- 
almost the entire set of levels contained in Table~\ref{PottsHeis.Table} are found below the filled symbols with
no extra level appearing. Thus we believe the combined spatial$+$spin symmetry pattern of the spectra 
for small magnetizations is a characteristic fingerprint of the 120$^\circ$ semiclassical states.

\textit{Finite-field case}.---
Here the ground state manifold is generally larger since the classical 
constraint $\vec{S}_\Delta=\vec{H}/2J$ allows for non-coplanar configurations already at the
level of a single triangle. It is known~\cite{Shender,ZhitoPRL02,Hassan} however that the coplanar states
with the spins lying in the field-plane have the largest number of soft modes and thus must be selected
by quantum or thermal fluctuations. A subsequent selection which depends on the field takes place within
the field-plane for the orientation of the spin triad~\cite{ZhitoPRL02}.
It turns out that the most relevant GS's in a finite field are the ones depicted in Figs.~\ref{CoplanarStates.Fig}(b)
and (c).~\footnote{Note that the remaining degeneracy is lifted already by harmonic spin waves~\cite{Hassan} 
with a selection of the $q=0$ state at small fields and a peculiar competition between the $q=0$ and the 
$\sqrt{3}\times\sqrt{3}$ state at higher fields (cf. Fig.~8 of Ref.~\onlinecite{Hassan}).} 
For $H < H_s/3$ (cf. Fig.~\ref{CoplanarStates.Fig}(b)) the relevant GS's on a triangle form a one-parameter family 
with one spin anti-parallel to the field. 
It is important to note that, apart from their difference in the directions of the three spins,
these configurations are spatially indistinguishable from the 120$^\circ$ states
of Fig.~\ref{CoplanarStates.Fig}(a): They are both 3-sublattice states and thus carry the same spatial
multiplicity (i.e. 24 for the cuboctahedron and 60 for the icosidodecahedron) 
and the same spatial symmetry (i.e., Eqs.~(\ref{CubocRed.Eq}) and (\ref{IcosiRed.Eq})).
Quite similarly, the ``quasi-collinear'' configurations of Fig.~\ref{CoplanarStates.Fig}(c), 
which can be selected for $H \geq H_s/3$, 
have an ``uud'' spatial structure. Hence they share the same spatial multiplicity and symmetry properties 
with the $M=M_s/3$ Ising GS's. Namely, there exist $9$ and $36$ ``quasi-collinear'' states for the cuboctahedron and the
icosidodecahedron respectively, and their spatial symmetry is given already in Eqs.~(\ref{CubocIsingRed.Eq}) 
and (\ref{IcosiIsingRed.Eq}).
We should note here that the set of IR's appearing in Eqs.~(\ref{CubocIsingRed.Eq}) and (\ref{IcosiIsingRed.Eq}) 
form a subset of the ones appearing in Eqs.~(\ref{CubocRed.Eq}) and (\ref{IcosiRed.Eq}) respectively.~\footnote{
An explanation of this feature can be readily given for the cuboctahedron: 
Here the 9 ``uud'' colorings can arise from the 24 vertex three-colorings $\mathbf{P}_{\text{ABC}}$ by identifying 
e.g. A with B and thus the symmetry IR's of the former are contained in the latter. 
Things are slightly different for the icosidodecahedron: Here the 60 vertex three-colorings of $\mathbf{R}_{\text{ABC}}$ 
provide (by identifying A with B) only the 30 ``uud'' colorings of $\mathbf{R}_{\text{uud}}^{(30)}$. 
Each of the remaining 6 ``uud'' states of $\mathbf{R}_{\text{uud}}^{(6)}$ contain two pentagonal loops of the type of 
Fig.~\ref{DiagPentagon.Fig}(a) (i.e. they have $n_a=2$) with five spins pointing up and thus cannot arise from a vertex 
three-coloring (since we cannot put alternating A, B spins on a pentagon).} 

According to the above, all types of semiclassical configurations of Fig.~\ref{CoplanarStates.Fig} have the same set 
of spatial representations, except the ``quasi-collinear'' states which do not contain the A2g representation: 
At large magnetizations this level is pushed higher in energy and this seems to confirm that it does not belong 
to the relevant towers of states of the ``quasi-collinear'' classical states. 
More generally, the fact that the same set of spatial IR's appear in the low-energy spectra at all magnetizations 
signifies that our previous semiclassical interpretation for the zero-field case carries over for finite fields as well. 
Again, one may ask for more stringent evidence by a comparison of the combined spatial$+$spin symmetry properties 
of these finite-magnetization states. We have derived these combined symmetries (not shown here) 
following the same lines as in Appendix~\ref{Tower.App}, but it turns out that the small size of the cuboctahedron 
together with the severe splitting of the low-lying states do not allow for a straightforward and thus definite 
identification of the relevant towers of states as above. As we show below, this will be possible for the much simpler 
case of the $s>1/2$ XY model.
 
\subsection{$s>1/2$ XY model}\label{XY.Subsec}
Here we study the XY model (i.e. the $J_z=0$ limit of Eqs.~(\ref{XXZ.Eq})-(\ref{Hxy.Eq}))
on the cuboctahedron. The reason of doing this is twofold. First because, in contrast to the Heisenberg point, 
the XY point exemplifies very evidently the core idea of the simultaneous appearance of several towers of states 
due to the spatial degeneracy of the classical GS's. And second, to provide an additional interpretation of 
the Heisenberg spectra, since these can be thought of as being adiabatically connected to the XY spectra but split 
by the quantum fluctuations introduced by $J_z$. 

%%%%%%%%%%%%%%%%%%%%%%%%%%%%%
\begin{figure}[!t]
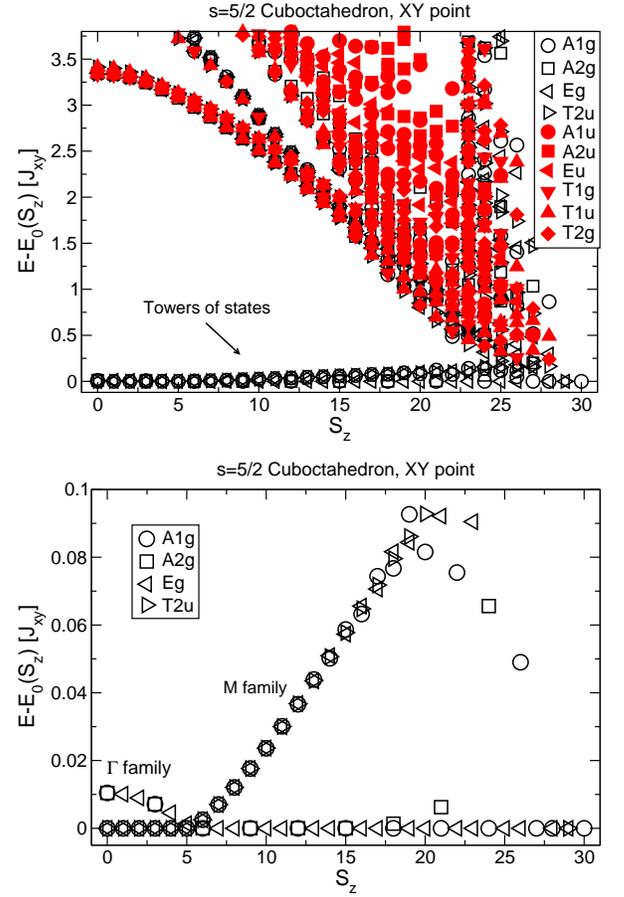

\centering
\includegraphics*[width=0.9\linewidth]{CubocFiveHalf_XY1}\\
\vspace{2mm}
\includegraphics*[width=0.9\linewidth]{CubocFiveHalf_XY1_Zoom}
\caption{(Color online) Low-energy spectra of the XY model on the $s=5/2$ cuboctahedron.
The arrow in the upper panel indicates the set of Anderson towers of states (or rotational bands).
These are shown in finer energy resolution in the lower panel in order to reveal their symmetry
structure. The latter is in full agreement with  Table\ref{PottsXY.Table}.
Note the level crossing between the two semiclassical families occurring slightly below the $M=M_s/3$ field.}
\label{Cuboc_XY.Fig}
\end{figure}
%%%%%%%%%%%%%%%%%%%%%%%%%%%%%

We first consider the classical case with a magnetic field perpendicular to the $xy$-plane.
As above, the XY Hamiltonian can be rewritten in terms of the three spins $\vec{s}_i$ of each triangle $\Delta$
and its total spin $\vec{S}_\Delta$ as (in units of $g\mu_B=1$):
\begin{eqnarray}\label{XYclassical.Eq}
\mathcal{H}^\mathrm{XY}_\mathrm{classical}=\frac{J_{xy}}{2}\sum_\Delta \Big( 
\mathbf{S}_{\Delta \perp}^2 -\sum_{i=1}^3 \big( \vec{s}_{i \perp}^2 + \frac{2 H}{J_{xy}} s_{i z}\big)\Big) ,
\end{eqnarray}
where $\mathbf{S}_{\Delta \perp}^2 \equiv \mathbf{S}_{\Delta x}^2+\mathbf{S}_{\Delta y}^2$ and similarly for 
$\vec{s}_{i \perp}^2$. The leading term of Eq.~(\ref{XYclassical.Eq}) is minimized by taking $\vec{S}_{\Delta \perp}=0$ 
on each triangle. The remaining terms require that we maximize both $\vec{s}_{i \perp}^2$ and $s_{i z}$
with the constraint $\vec{s}_{i \perp}^2+s_{i z}^2 = s(s+1)$ (the balance between the two components is controlled 
by the ratio $2H/J_{xy}$). In zero-field this gives the 3-sublattice states where all triangles share the 
same spin (xy) plane. The major difference with our previous analysis of the Heisenberg (classical) GS's is that
here a spin plane (i.e. the $xy$ plane) is selected explicitly from the beginning (i.e., for zero field)
and this gives a finite energy cost to non-coplanar configurations. 
A finite field gives rise to a tilt of the three spins out of the xy plane giving rise to the so-called 
``umbrella'' states. It is clear that both in zero and in finite field, the classical XY GS's have the same spatial
multiplicity and spatial symmetry properties (but not spin symmetry properties, see below)
with the set of coplanar 3-sublattice GS's of the Heisenberg point.

%%%%%%%%%%%%%%%%%%%%%%%%%%%%%
\begin{table}[!t]
\caption{XY point: Decomposition of semiclassical coplanar states into IR's of
$\mathsf{G}=\mathsf{C}_{\infty v}\times \mathsf{R}$ (where $\mathsf{R}=\mathsf{O}_h$ or $\mathsf{I}_h$).
N is the number of IR's for each sector of $\mathsf{C}_{\infty v}$ and is equal to the dimensionality
of the corresponding IR of $\mathsf{C}_{\infty v}$
(cf.Table~\ref{Cv.Table}) times the number of coplanar states in each family divided by six
(i.e., 1 for $\mathbf{P}^\Gamma_{\text{ABC}}$, 3 for $\mathbf{P}^M_{\text{ABC}}$, and 10 for $\mathbf{R}_{\text{ABC}}$).
The details of the derivation are given in Appendix~\ref{Tower.App}.
\label{PottsXY.Table}}
\begin{ruledtabular}
\begin{tabular}{l|lr|lr|lr}
                 &Cuboc.& &Cuboc. & &Icosi. \\
$(S_z, \sigma_v)$&$\mathbf{P}^\Gamma_{\text{ABC}}(1\times 6)$& N& $\mathbf{P}^M_{\text{ABC}}(3\times 6)$& N &$\mathbf{R}_{\text{ABC}}(10\times 6)$ & N\\
\hline
$0, +$           &A1g     &1 &A1g,Eg     & 3 & Ag,Au,Fg,Fu     &10\\
$0, -$           &A2g     &1 &T2u        & 3 & Ag,Au,Fg,Fu     &10\\
\hline
$1,2,4,.$        &Eg      &2 &A1g,Eg,T2u & 6  & 2Hg,2Hu     &20\\
\hline
$3,6,9,.$        &A1g,A2g &2 &A1g,Eg,T2u & 6  & 2(Ag,Au,Fg,Fu) &20
\end{tabular}
\end{ruledtabular}
\end{table}
%%%%%%%%%%%%%%%%%%%%%%%%%%%%%
%%%%%%%%%%%%%%%%%%%%%%%%%%%%%%
%\begin{table}[!t]
%\caption{IR's of $\mathsf{G}=\mathsf{U}(1)\times \mathsf{R}$
%(where $\mathsf{R}=\mathsf{O}_h$ or $\mathsf{I}_h$) contained in the Anderson tower of states
%of the XY model. Note that the IR's in the icosidodecahedron case
%should be counted twice. For the derivation see Appendix~\ref{Tower.App}.\label{PottsXY.Table}}
%\begin{ruledtabular}
%\begin{tabular}{l|ll|l}
%           &  Cuboc.              &                              &  Icosi.  \\
%$S_z$      & $\mathbf{P}_\Gamma (1\times 6)$& $\mathbf{P}_M (3\times 6)$   & $\mathbf{P'} (10\times 3)$\\
%\hline
%$0,3,6,.$  & A1g,A2g              &A1g,Eg,T2u  & Ag,Au,Fg,Fu  \\
%$1,2,4,.$  & Eg                   &A1g,Eg,T2u  & Hg,Hu       \\
%\end{tabular}
%\end{ruledtabular}
%\end{table}
%%%%%%%%%%%%%%%%%%%%%%%%%%%%%

Let us now consider the quantum-mechanical XY model. For our demonstration purposes, it suffices to consider the $s=5/2$ 
case only (the $s=3/2$ case is very similar\footnote{Interestingly the $s=1/2$ XY model spectrum (not shown here) 
resembles much more the $s=1/2$ Heisenberg spectra of Fig.~\ref{Cuboc_Heis.Fig}, and does {\em not} show the well separated 
tower of states of the large $s$ XY model.}). The low-energy spectrum is shown in the upper panel of Fig.~\ref{Cuboc_XY.Fig}.
A number of spectral features are revealed. First, in contrast to the Heisenberg case studied above, the lowest-energy 
portion of the spectrum (indicated by the arrow) is well isolated from higher excitations and it comprises 
four {\em distinct} towers of states. This multiplicity is a fingerprint of the spatial degeneracy of the classical 
3-sublattice states mentioned above.
This can be further substantiated by examining more closely the symmetry structure of the excitations intervening in these
towers. To this end we zoom in on these towers in the lower panel of Fig.~\ref{Cuboc_XY.Fig}.
A simple inspection of the spatial IR's that appear in this panel reveals that they are exactly the ones given 
by Eq.~(\ref{CubocRed.Eq}). 
%Although this is already an indication that the low-energy excitations are of semiclassical 
%origin it does not however exclude at all the possibility that a decomposition of some other types of classical states 
%can contain some (or all) of the IR's of Eq.~(\ref{CubocRed.Eq}).
Much stronger evidence comes by examining the full spatial$+$spin symmetry structure of the lowest towers. 
Indeed, a closer comparison to Table~\ref{PottsXY.Table} (derived in Appendix~\ref{Tower.App}) 
demonstrates that there is a remarkable one-to-one correspondence of each of the lowest towers with the classical 
families which holds in almost the entire magnetization range. 
We should emphasize here that each semiclassical state shows a different and quite non-trivial 
symmetry pattern which is in some sense a very characteristic fingerprint of the state. For instance, the full content 
of 24 states of Eq.~(\ref{CubocRed.Eq}) is recovered every three $S_z$ sectors
in the lowest towers with the specific pattern of combined spatial and spin IR's given in Table~\ref{PottsXY.Table}.
This remarkable agreement between exact ED spectra and our symmetry derivation is indeed a strong evidence
that the lowest towers of states can be thought of as renormalized semiclassical 3-sublattice states.

Some additional remarks are in order here regarding the energies of the two families of towers as revealed in the lower panel of
Fig.~\ref{Cuboc_XY.Fig}. We should first note that all towers are expected to become degenerate in the classical
$s\gg 1$ limit. Based on spectra with $s=3/2$ (not shown here), we find that the energy splittings within
a tower of each of the two families $\mathbf{P}^\Gamma_{ABC}$ and $\mathbf{P}^M_{\text{ABC}}$ diminishes
quickly on increasing $s$ as expected, while at the same time, the energy splitting between them remains sizable.
On the other hand, as a function of $S_z$ (or field), there is an interesting level crossing between the two families 
somewhat below the $M=M_s/3$ magnetization plateau, with the $M$ family being more favorable below this point. 
An understanding of the above level-crossing could arise by employing for instance a semiclassical expansion
for the XY model in a field, in a similar fashion with what is done for the Heisenberg model~\cite{Henley,Hassan},
but such an analysis is clearly beyond the scope of this paper.

\section{Summary}\label{Discussion.Sec}
We have presented an extended study of the low-energy physics of two existing magnetic molecule
realizations of the kagom\'e AFM on the sphere, the cuboctahedron and the icosidodecahedron.
Our ED results revealed a number of generic spectral features which stem from the corner-sharing topology
of these clusters. Indeed, a simple comparison to a finite-size $s=1/2$ unfrustrated magnet demonstrated that
frustrated clusters manifest a ``bulk'' of very dense low-energy excitations. We focused on two major aspects 
which are of general interest but were particularly oriented toward 
the $s=5/2$ Mo$_{72}$Fe$_{30}$ cluster: (i) the low-energy excitations above 
the $M=M_s/3$ plateau and (ii) the low-lying spectra of the Heisenberg model for $s>1/2$. 

For the $M=M_s/3$ plateau, we first demonstrated that the $s=1/2$ icosidodecahedron shows 36 low-lying 
excitations which are adiabatically connected to collinear ``uud'' Ising (GS's), at the same time being well isolated
from higher levels by a relatively large energy gap. We then argued, based on a complementary physical picture 
which emerged from the derivation of an effective quantum dimer model, that this $s=1/2$ feature must be special 
to the topology of the icosidodecahedron and that it must survive for $s=5/2$ as well. We also predicted that 
the corresponding 36 low-lying plateau states of the $s=5/2$ icosidodecahedron consist of two ``uud'' families (of 30 and 6 states respectively)   
which are separated by a small diagonal energy splitting. This result can be confirmed by low-temperature specific 
heat measurements at the $M=M_s/3$ regime ($H\simeq 5.9$ Tesla) of Mo$_{72}$Fe$_{30}$ 
and/or by an assessment of the associated missing entropy. 

In the second part, we showed exact diagonalization spectra for the $s>1/2$ Heisenberg cuboctahedron which 
demonstrated that the dense low-lying excitation features of the $s=1/2$ case are present for $s>1/2$ as well, albeit with 
a striking spatial$+$spin symmetry pattern. 
These spectra provide a semiclassical interpretation of the broad inelastic neutron scattering response reported 
for Mo$_{72}$Fe$_{30}$. The main ingredient of this interpretation is the simultaneous presence of several low-energy 
towers of states or rotational bands at low energies which originate from the large spatial multiplicity 
of the classical Heisenberg ground states, and this is known to be a generic feature of highly frustrated clusters. 
This semiclassical interpretation was further corroborated by an independent group theoretical analysis 
which demonstrated that the striking symmetry pattern of the low-lying excitations is indeed 
a characteristic fingerprint of the classical {\em coplanar} ground states. 
The core idea of the simultaneous presence of several rotational bands at low energies 
due to the discrete classical degeneracy was finally exemplified very evidently by a study of the $s>1/2$ XY model.

\section{Acknowledgments}
We would like to acknowledge fruitful discussions with C. L. Henley, M. Luban, and K. P. Schmidt
and earlier collaboration with J.-B. Fouet and S. Dommange on related subjects.
The point group symmetry data used in our calculations have been obtained using the
Bethe package~\cite{Bethe}. We would like to thank K. Rykhlinskaia for her assistance
with this package. This work was supported by the Swiss National Fund and by MaNEP.
The computations have been enabled by the allocation of computational
resources on the machines of the CSCS in Manno.

\appendix
\section{Degenerate Perturbation Theory around the Ising point}\label{DPT.App}
Here we describe some very general considerations which greatly facilitate the classification
of processes appearing in degenerate perturbation theory around the Ising point.
We consider corner sharing triangles structures with general spin $s$, and focus on the $M=M_s/3$ plateau.
The ground-state Ising manifold consists of configurations which maximize the number of
extremum local moments $m$~\cite{Bergman1}. At the plateau $M=M_s/3$ phase, this consists of all configurations
with two spins having $m=s$ and one with $m=-s$ in each triangle. All processes triggered by $\mc{H}_{xy}$
must preserve this constraint. Let us denote by $E_0$ the zero-th order energy and by
$\mc{P}_0,\mc{Q}_0=1-\mc{P}_0$ the projections onto and out of the Ising manifold.
We also designate by $\mc{R}$ the resolvent operator
\be
\mc{R} =(E_0-\mc{Q}_0\mc{H}_0\mc{Q}_0)^{-1}~.
\ee
The $n$-th order term of the effective Hamiltonian in the Rayleigh-Schr\"odinger formulation\cite{DPT} reads
\be\label{DPT.Eq}
\mc{H}_{\text{eff}}^{(n)}=\mc{P}_0\mc{H}_{xy}(\mc{R}\mc{H}_{xy})^{n-1}\mc{P}_0 + ~\text{remaining terms}~,
\ee
where each of the ``remaining terms'' can be thought of as a product combination of lower order terms
$\mc{H}_{\text{eff}}^{(k)}$ (with $k < n$). This separation is useful when e.g. all processes below some
order $n$ are constant since then, to order $n$, it suffices to keep only the leading term of
Eq.~(\ref{DPT.Eq}). The presence of $\mc{P}_0$'s in Eq.~(\ref{DPT.Eq}) enforces all terms to flip spins
in such a way as to respect the ``uud'' constraint in each triangular unit.

The derivation of $\mc{H}_{\text{eff}}$ at any given order is greatly facilitated by noting that only ``linked''
processes should be taken into account. These are interactions that are ``connected'' in the following sense.
Substituting $\mc{H}_{xy}$ in Eq.~(\ref{DPT.Eq}) gives different types of terms, each one carrying a string
of a given number of bond operators $s_i^-s_j^+$. We differentiate between ``linked'' and ``unlinked''
interaction terms depending on whether the set of all vertices appearing in the corresponding string forms a
connected (open or closed) path in the lattice or not. The latter contain nonlocal interactions
between disconnected parts of the lattice and therefore must be omitted~\cite{DPT}.
Only connected paths should therefore be considered.
Further simplifications arise from the ``uud'' constraint as described below for diagonal and off-diagonal
processes.
\begin{figure}
\centering
\includegraphics*[width=0.7\linewidth]{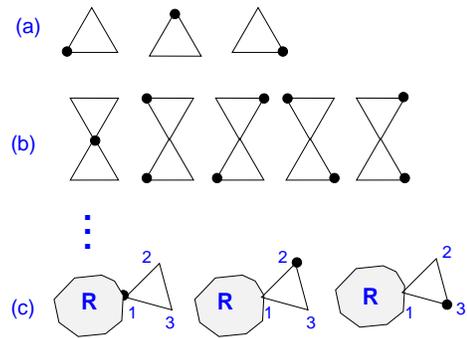}
\caption{As above, filled circles denote spins with $m=-s$, remaining vertices have $m=s$.
(a) All configurations of a single triangle have the same energy to all orders.
(b) The configurational energy of two adjacent triangles depends only on the shared spin, since all
4 states with the shared spin pointing up have the same energy.
(c) The contraction method is based on the observation that $E(R;s_1,s_2,s_3) = E(R;s_1)$.}
\label{Contractions.Fig}
\end{figure}

\textit{Diagonal processes}.---
As explained in considerable detail in Ref.~\onlinecite{Bergman1} in the context of the pyrochlore lattice,
all diagonal processes up to a given order give an overall constant energy shift, with the leading non constant 
terms arising from processes along closed loop configurations. As we explain below, similar results apply to the 
present case of corner sharing triangles as well. The proof can be demonstrated in a compact way 
by using the contraction method of Bergman \textit{et al.}~\cite{Bergman1}

Consider first all-order diagonal processes confined to a given triangle.
Clearly, the only physically distinct configuration on this triangle is the ``uud'' one, since
the associated diagonal energy does not depend on which of the three vertices the down spin resides
(cf. Fig.~\ref{Contractions.Fig}(a)).
By sampling the energies of all triangles we end up with an energy that is global, i.e., the same for all Ising states. 
Hence all-order processes confined in a single triangle give an overall constant energy shift
and can thus be neglected. We now consider all-order processes confined to two adjacent triangles.
Here, there are two physically distinct classes of configurations (cf. Fig.~\ref{Contractions.Fig}(b)),
depending on whether the shared spin points down (first state in Fig.~\ref{Contractions.Fig}(b)) or up (remaining four states).
Since all four states with the shared spin pointing up are physically equivalent, the energy is only a function of the shared 
spin variable. By sampling again over the lattice we obtain a global energy shift. Remarkably, these arguments can be generalized to much broader
cases by the contraction method exemplified in Fig.~\ref{Contractions.Fig}(c):
Since permuting the spin variables 2 and 3 results in a topologically equivalent
state, knowing the value of $s_1$ is enough, i.e.,
\be
E(R;s_1,s_2,s_3) = E(R;s_1)~,
\ee
where $R$ designates all remaining spin variables. This ``contraction'' can be continued with the next
available triangle inside the shaded area $R$, and so on. If this process can be continued until we are left
with a function of a single vertex, then sampling this function over all spins we obtain
again a global energy shift.
\begin{figure}%[!t]
\centering
\includegraphics*[width=0.5\linewidth]{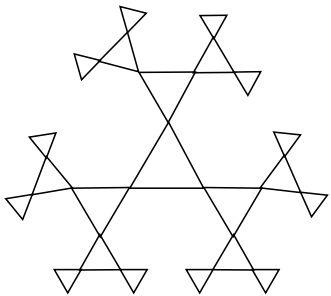}
\caption{A fragment of the corner-sharing triangle lattice with no closed loops. This is a Bethe lattice
made of triangles, known as Husimi cactus.}\label{HusimiCactus.Fig}
\end{figure}
A non-constant energy contribution may arise only when the contraction process cannot be continued until the
last spin variable. This happens whenever the shaded area $R$ of Fig.~\ref{Contractions.Fig}(c) contains one or
more closed loops, since none of the triangles making up a loop is ``contractible''. This leaves us with the
following quite general statement: All-order processes confined to a fragment of the lattice
with no closed loops give an overall constant energy shift. The most general form of such fragments
is depicted in Fig.~\ref{HusimiCactus.Fig} and is recognized to be a Bethe lattice made of
triangles, known as Husimi cactus.

Hence, the lowest order diagonal processes come from closed loops in the lattice.
One should also remark that, in contrast to off-diagonal processes (see below),
the order at which diagonal processes first appear is independent of the spin $s$.

\begin{figure}%[!t]
\centering
\includegraphics*[width=0.5\linewidth]{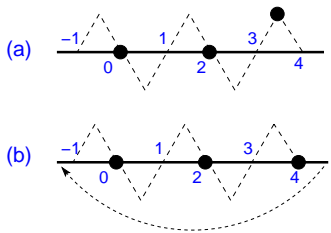}
\caption{The thick solid line denotes a path corresponding to a given term in the perturbation
series. Only the alternating spin up-down configuration shown in (b) (which must be closed)
is amenable to an off-diagonal kinetic process.}
\label{OffDiagPaths.Fig}
\end{figure}
\begin{table}%[!tbh]
\caption{Degenerate perturbation theory calculations for the off-diagonal kinetic amplitude $t$ on
alternating up-down configurations around loops with $L=4$, $6$ and $10$ sites,
and for various intrinsic spins $s$. The order in $\mc{H}_{xy}$ and the total number of
contributing processes are also given.}
\label{Amplitudes.Table}
\begin{ruledtabular}
\begin{tabular}{ccclr}
L & 2s & Order ($L s$)&  $t~[J_z\alpha^{L s}]$ & \# of processes\\
\hline
4&1 & 2 & -1.0       & 4  \\
 &2 & 4 & -1.0       & 36 \\
 &3 & 6 &  -0.5625   & 400\\
 &4 & 8 & -0.25      & 4 900\\
 &5 & 10& -0.09765625& 63 504\\
 &6 & 12& -0.03515625& 853 776\\
\hline
6&1 & 3 &  1.5         & 12  \\
 &2 & 6 & -0.88402469  & 900 \\
 &3 & 9 &  0.25093125  & 94 080\\
 &4 & 12& -0.05637473  & 11 988 900\\
 &5 & 15&  0.01106939  & 1 704 214 512 \\
 &6 & 18& -0.00199964  & 260 453 217 024\\
\hline
8&1 &  4 &-2.5& 48\\
 &2 &  8 &-0.93709194 &45 360 \\
 &3 & 12 &-0.12770306 &60 614 400\\
 &4 & 16 &-0.01464521 &114 144 030 000 \\
\hline
10&1 & 5 &+4.375 & 240\\
  &2 & 10 &-1.0924858& 3 855 600
\end{tabular}
\end{ruledtabular}
\end{table}

\textit{Off-diagonal processes}.---
Figure \ref{OffDiagPaths.Fig} shows schematically a given lattice path (solid thick line)
with two particular configuration choices.
All spins have $m=s$ except the ones indicated by filled circles with $m=-s$.
By definition, only spins residing on this path may be flipped.
In off-diagonal processes at least one spin, say $\vec{s}_0$, is flipped from $m=\pm s$
to $m=\mp s$. But, since the final state must preserve the ``uud'' constraint in each triangle,
flipping $\vec{s}_0$ must be accommodated by a similar flip of the adjacent spins lying on the path, namely
$\vec{s}_{-1}$ and $\vec{s}_1$. Similarly, the spins at vertices 2 and 3 must also be flipped.
If, as drawn in Fig.~\ref{OffDiagPaths.Fig}(a), the initial configuration has $m_3=m_4=-s$, then flipping $3$
already violates the ``uud'' constraint.
It is also clear that if the path is open at one end (or at both ends) the ending spin (or spins)
will be flipped at the final state, thus violating the ``uud'' constraint on the adjacent triangle(s).
Thus only closed, connected paths with alternating up-down spins, such as in Fig.~\ref{OffDiagPaths.Fig}(b)
are amenable to an off-diagonal process.
The simplest possible paths are loops of even number of spins $L$. Since flipping each spin requires $2 s$
operations, off-diagonal processes on a simple loop appear in order $\alpha^{L s}$.
An example was shown in Fig.~\ref{IcosiOffDiagLoop.Fig} for the icosidodecahedron.
In Table \ref{Amplitudes.Table} we provide calculated numerical values for the amplitude of such virtual processes
around loops with $L=4$, $6$, $8$ and $10$ sites and various values of $s$.
Starting from a given loop, one may also build processes (and paths) of higher order by invoking adjacent
triangular units. An example for the icosidodecahedron was the fifth order processes mentioned
in Sec.~\ref{Icosi_Plateau.Subsubsec}.

\section{Symmetry properties of the 3-sublattice coplanar states}\label{Tower.App}
Here we give the details of the derivation of the symmetry properties of the semiclassical 3-sublattice coplanar states
of Fig.~\ref{CoplanarStates.Fig}(a). These are relevant for the Heisenberg model at $H=0$ as well as the XY model 
(discussed in Sec.~\ref{XY.Subsec}) for both zero and finite fields. 
As explained above, the spatial symmetry properties of the states of Fig.~\ref{CoplanarStates.Fig}(b) are the 
same as that of (a), and similarly the spatial symmetry of the ``quasi-collinear'' states of Fig.~\ref{CoplanarStates.Fig}(c) 
is identical to that of the ``uud'' states given in Eqs.~(\ref{CubocIsingRed.Eq}) and (\ref{IcosiIsingRed.Eq}). 
The derivation of the combined spatial$+$spin properties of Fig.~\ref{CoplanarStates.Fig}(b) and (c) are not of 
our interest here but can be found easily following the same steps as below.

We are using the following notation and conventions. The groups of real space and spin space operations 
are denoted respectively by $\mathsf{R}$ and $\mathsf{L}$. In particular, $\mathsf{R}=\mathsf{O}_h$ for the cuboctahedron
and $\mathsf{I}_h$ for the icosidodecahedron. The full group $\mathsf{R}\times\mathsf{L}$ is designated by $\mathsf{G}$.
A stabilizer $\mathsf{H}_c$ of a classical state $|c\rangle$ consists of elements $h$ which preserve $|c\rangle$ 
(i.e. $h|c\rangle=|c\rangle$) and this will be either a subgroup of $\mathsf{R}$ or a subgroup of $\mathsf{G}$ 
depending on whether we are examining only the spatial or the full symmetry properties.
The elements of $\mathsf{R}$, $\mathsf{L}$ and $\mathsf{G}$ are labeled by $r$, $l$, $g=r\cdot l$ respectively, 
while their IR's are denoted as $\mathbf{D}^\rho(r)$, $\mathbf{D}^\lambda(l)$, and 
$\mathbf{D}^\gamma(g) = \mathbf{D}^\rho(r)\otimes\mathbf{D}^\lambda(l)$.
Similarly, their characters are denoted by $\chi^\rho(r)$, $\chi^\lambda(l)$, and 
$\chi^\gamma(g=r\cdot l)=\chi^\rho(r)\cdot\chi^\lambda(l)$.
Let us first discuss the spatial symmetry and then the combined spatial$+$spin symmetry structure of the 3-sublattice 
states.
 
\subsection{Spatial symmetry of 3-sublattice states}
As we discussed previously, each vertex three-coloring $|c\rangle$ is in one-to-one correspondence with the states 
of Fig.~\ref{CoplanarStates.Fig}(a). 
Starting from a given $|c\rangle$ and applying all elements $r$ of $\mathsf{R}$ we generate an invariant vector space 
or orbit $\mathbf{O}$. The decomposition of $\mathbf{O}$ into IR's $\mathbf{D}^{(\rho)}$ of $\mathsf{R}$ is given 
by the well known formula~\cite{GroupTheory}
\begin{eqnarray}
\mathbf{O}(r) &=& \sum_\rho  m_\rho \mathbf{D}^{\rho}(r)~,\label{SpatialDecomp.Eq}\\
m_\rho&=& \frac{1}{|\mathsf{R}|} \sum_{r \in \mathsf{R}} \chi^{\rho}(r)^*
~\text{Tr}[\mathbf{O}(r)]~.\label{mrho1.Eq}
\end{eqnarray}
The matrix element $\mathrm{O}_{cc'}(r)=\langle c|r|c'\rangle$ is equal to one if $r$ belongs to the stabilizer 
$\mathsf{H}_c$ of $|c\rangle$ and vanishes otherwise. Thus we may rewrite Eq.~(\ref{mrho1.Eq}) as
\be\label{mrho2.Eq}
m_\rho = \frac{1}{|\mathsf{H}_c|} \sum_{h \in \mathsf{H}_c\subseteq \mathsf{R}} \chi^{\rho}(h)~,
\ee
where we made use of $|\mathsf{R}|=|\mathbf{O}|\cdot |\mathsf{H}_c|$. This relation follows from the coset 
decomposition of $\mathsf{R}$ with respect to $\mathsf{H}_c$ (each coset is in one-to-one correspondence 
with the states $|c'\rangle$ of the orbit $\mathbf{O}$). Employing (\ref{mrho2.Eq}) to all different orbits 
we obtain the symmetry properties of the coplanar states.

Let us see now what happens for the cuboctahedron and the icosidodecahedron separately.
As discussed above, the cuboctahedron has a total number of 24 vertex three-colorings $|c\rangle=|\text{ABC}\rangle$.
Under $\mathsf{O}_h$ they form four invariant orbits of six colorings each.
The first orbit, called $\mathbf{P}^\Gamma_{\text{ABC}}$ consists of the six global permutations
of the translationally invariant coloring depicted in Fig.~\ref{CubocOnThePlane.Fig}(a).
The 18 colorings that belong to the remaining three orbits result from the first orbit by interchanging
colors along loops with two alternating colors. One such configuration is shown in Fig.~\ref{CubocOnThePlane.Fig}(b).
Although these three orbits are equivalent it is useful for the following discussion of the full spatial$+$spin properties 
to treat them collectively as a single one which we term $\mathbf{P}^M_{\text{ABC}}$.~\footnote{Our notation for these two 
orbits is borrowed from the corresponding points of the Brillouin zone of the 12-site kagom\'e, i.e. the $k=0$ point 
and the three $M$-momenta at the middle points of the boundary edges. Note however that the three orbits of the M family 
are not in one-to-one correspondence with the three $M$-momenta but rather combine all of them.}
The decomposition of the above orbits into IR's of $\mathsf{O}_h$ was given in Eq.~(\ref{CubocRed.Eq}).

On the other hand the icosidodecahedron has 60 coloring states. Under $\mathsf{I}_h$ they form two
orbits of 30 colorings each. The first orbit consists of the 3 cyclic permutations (ABC,CAB,BCA)
of 10 ABC states, while the second orbit consists of the remaining 3 permutations (or reflections) (BAC,CBA,ACB)
of the same 10 states. Although these two orbits are equivalent it is useful for the discussion of the full
spatial$+$spin properties (see below) to treat them as a single one which we denote by $\mathbf{R}_{\text{ABC}}$.
Its decomposition into IR's of $\mathsf{I}_h$ was given in Eq.~(\ref{IcosiRed.Eq}).

\subsection{Spatial$+$spin symmetry of 3-sublattice states}
We shall now go one step further and derive the combined spatial$+$spin properties of the above states.
Namely, a decomposition similar to that of Eqs.~(\ref{CubocRed.Eq}) and (\ref{IcosiRed.Eq}) but now in terms of 
IR's $\mathbf{D}^{\gamma}$ of the full symmetry group $\mathsf{G}=\mathsf{R}\times\mathsf{L}$ of the Hamiltonian.
The method has been employed previously in the seminal works of Bernu \textit{et al.}~\cite{Bernu1,Bernu2} 
and Lecheminant \textit{et al.}~\cite{Lecheminant_J1J2,Lecheminant_kagome}.

The recipe is quite analogous to the one we employed above. Here however, by applying the elements of the full group 
$\mathsf{G}$ on a given classical state we generate a continuous orbit $\mathbf{O}$.
All states contained in this orbit are coplanar but their order parameter has all possible orientations in spin space
($\mathbf{O}$ is a continuous ``order parameter space''). Otherwise, the equation giving the numbers $m_\gamma$ 
(i.e. how many times is $\mathbf{D}^{\gamma}$ appearing in the decomposition of $\mathbf{O}$ into IR's of $\mathsf{G}$) 
is fully analogous to Eq.~(\ref{mrho2.Eq}) and reads
\be\label{mgamma.Eq}
m_\gamma=\frac{1}{|\mathsf{H}_c|} \sum_{h \in \mathsf{H}_c\subseteq \mathsf{G}} \chi^{\gamma} (h),
\ee
where now $\chi^{\gamma}(h=r\cdot l)=\chi^{\rho}(r)\cdot\chi^{\lambda}(l)$.
For the identification of $\mathsf{H}_c$ it is expedient to split the operations of the space group
$\mathsf{R}$ into the $3!$ sets $\mathsf{S}_{abc},\mathsf{S}_{cab},\cdots,\mathsf{S}_{cba}$
defined as follows. The first consists of elements which map $|c\rangle$ to itself. On the other hand,
the elements of $\mathsf{S}_{cab}$ map $|c\rangle$ to its globally permuted (ABC) $\mapsto$ (CAB) version,
and similarly for the remaining sets. We also define the following set of integer numbers
\be
N_{abc}^{\rho}=\sum_{h\in \mathsf{S}_{abc}} \chi^{\rho}(h),~
N_{cab}^{\rho}=\sum_{h\in \mathsf{S}_{cab}} \chi^{\rho}(h),~\text{etc}~.
\ee
These numbers depend on the transformation properties of $|c\rangle$ under the spatial group $\mathsf{R}$ alone.
The non-vanishing ones are given in Table~\ref{Nvalues.Table}.
%%%%%%%%%%%%%%%%%%%%%%%%%%%%%
\begin{table}[!t]
\caption{Character table of $\mathsf{C}_{\infty v}$ (see e.g. Ref.~\onlinecite{GroupTheory}).
Here, $E$ denotes the identity, $R_\phi$ the set of $\mathsf{U}(1)$ rotations,
and $\sigma_v$ the set of all vertical mirror planes. The IR's $\mathsf{C}_{\infty v}$ are characterized
by $|S_z|$, and the ``parity'' under $\sigma_v$ for $S_z=0$.\label{Cv.Table}}
\begin{ruledtabular}
\begin{tabular}{l||ccc}
$|S_z|, \sigma_v$ & $E$ & $R_\phi$ & $\sigma_v$ \\
\hline
\hline
$0,+$ &  1 &           1&  1 \\
$0,-$ &  1 &           1&  -1 \\
$n\geq 1$ &  2 & 2 $\cos{n \phi}$&  0
\end{tabular}
\end{ruledtabular}
\end{table}
%%%%%%%%%%%%%%%%%%%%%%%%%%%%%

We should note here that one can make a choice of $\mathsf{G}$ depending on the amount of information
we seek. For instance we may choose according to the symmetries we implement in our exact diagonalizations.
We may even take $\mathsf{L}$ as the idenity, i.e. $\mathsf{G}=\mathsf{R}$. In the latter case we 
recover Eq.~(\ref{mrho2.Eq}).
The type and number of invariant vector spaces in each case will be different and the symmetry decomposition
must be applied to each one separately, but the corresponding results will be consistent with each other.

Let us now apply the above to the zero-field Heisenberg model and the XY model.

%%%%%%%%%%%%%%%%%%%%%%%%%%%%%%%%%%%
\begin{table}\caption{The numerical values of the integers $N_{abc}$, $N_{cab}$, etc.
defined in the text. Remaining IR's or blank entries correspond to vanishing values.
We also give the order of the corresponding stabilizers $\mathsf{H}_c \subseteq \mathsf{G}$.}
\label{Nvalues.Table}
\begin{ruledtabular}
\begin{tabular}{l|c|c|c}
         &  Cuboc. $\Gamma$&  Cuboc. $M$&   Icosi. \\
         &  (A1g,A2g,Eg)&    (A1g,Eg,T2u)& (Ag,Au,Fg,Fu,Hg,Hu)\\
\hline
\hline
$|\mathsf{H}_c|$ &48     &             16   &            12\\
\hline
$N_{abc}$&     (8, 8,16)&         (8, 8, 8)&     (4, 4, 4, 4, 8, 8)\\
$N_{cab}$&     (8, 8,-8)&                  &     (4, 4, 4, 4,-4,-4)\\
$N_{bca}$&     (8, 8,-8)&                  &     (4, 4, 4, 4,-4,-4)\\
$N_{bac}$&     (8,-8, 0)&         (8, 8,-8)&             \\
$N_{acb}$&     (8,-8, 0)&                  &             \\
$N_{cba}$&     (8,-8, 0)&                  &             \\
\end{tabular}
\end{ruledtabular}
\end{table}
%%%%%%%%%%%%%%%%%%%%%%%%%%%%%

\subsubsection{$\mathsf{SU}(2)$ point}
Here we take $\mathsf{L}=\mathsf{SU}(2)$ and $\lambda$ is the total spin $S$ which is integer here.
%~\footnote{The following 
%analysis applies for half-integer spins as well but ones must enlarge $\mathsf{L}$ to the so-called double group.}
A single $|c\rangle$ generates the full set of coplanar states for each family of the clusters.
The elements of $\mathsf{S}_{bac}$, $\mathsf{S}_{cba}$, and $\mathsf{S}_{acb}$
can be combined with $\pi$ rotations of $\mathsf{SU}(2)$ and bring $|c\rangle$ back to itself.
Thus again $|\mathsf{H}_c|=|\mathsf{S}_{abc}|+\cdots+|\mathsf{S}_{cba}|$.
Using Eq.~(\ref{mgamma.Eq}) with $\gamma=(\rho,S)$ then
\begin{eqnarray}\label{PottsHeis.Eq}
m_{\rho,S}&=& \Big[ N_{abc}^{\rho}\chi^{(S)}(0)+
N_{cab}^{\rho} \chi^{(S)}(\frac{2\pi}{3}) +
N_{bca}^{\rho} \chi^{(S)}(\frac{4\pi}{3}) \nonumber\\
&&+\left( N_{bac}^{\rho}+N_{acb}^{\rho}+N_{cba}^{\rho}\right) \chi^{(S)}(\pi)
 \Big] / |\mathsf{H}_c|~.
\end{eqnarray}
where~\cite{GroupTheory} $\chi^{(S)}(\phi)=\frac{\sin{(S+\frac{1}{2})\phi}}{\sin{\phi/2}}$.
Replacing the values of Table~\ref{Nvalues.Table} for each family separately
we obtain the symmetry structures given in Table \ref{PottsHeis.Table}.

\subsubsection{XY point}
Here, we take $\mathsf{L}=\mathsf{C}_{\infty v}$ which includes $\mathsf{U}(1)$ rotations and
the continuous set of vertical (i.e. containing the $z$-axis) mirror planes.
The IR's of $\mathsf{C}_{\infty v}$ can be generally labeled~\cite{GroupTheory}
by a non-negative integer $n$ or $|S_z|$, with an additional label $\sigma_v=\pm 1$ for $n=0$
which stands for the parity under the mirror operation. Hence $\lambda=(|S_z|,\sigma_v)$.
All IR's for $n=|S_z|\neq 0$ are two-dimensional and consist of pairs of $S_z$ and $-S_z$ basis vectors.
The characters of $\mathsf{C}_{\infty v}$ are given in Table~\ref{Cv.Table}.

It suffices to select a single coloring state $|c\rangle$ since this generates all coloring states
for both clusters. The stabilizer $\mathsf{H}_c$ can be found as follows.
Each one of the sets $\mathsf{S}_{abc}$, $\mathsf{S}_{cab}$, etc. discussed above can be combined with one of 
the elements of $\mathsf{C}_{\infty v}$ to give $|c\rangle$ again. For instance, an element of $\mathsf{S}_{cab}$ 
can be combined with a $\mathsf{U}(1)$ spin rotation of $2\pi/3$. On the other hand, an element of $\mathsf{S}_{bac}$
can be combined with a mirror plane containing the $C$-axis (i.e. that of the spins ``colored'' as $C$). The set of all such 
combined operations span $\mathsf{H}_c$, i.e. $|\mathsf{H}_c|=|\mathsf{S}_{abc}|+\ldots+|\mathsf{S}_{bac}|$. 
Using Eq.~(\ref{mgamma.Eq}) with $\gamma=(\rho,\lambda)$ then
\begin{eqnarray}\label{PottsXY.Eq}
m_{\rho,\lambda}&=&[N_{abc}^{\rho}\chi^{\lambda}(E)+
N_{cab}^{\rho}\chi^{\lambda}(\frac{2\pi}{3})+
N_{bca}^{\rho} \chi^{\lambda}(\frac{4\pi}{3})\nonumber\\
&&+(N_{bac}^{\rho}+N_{acb}^{\rho}+N_{cba}^{\rho})\chi^{\lambda}(\sigma_v) ]/|\mathsf{H}_c|~.
\end{eqnarray}
Replacing the values of Table~\ref{Nvalues.Table} and the characters $\chi^{\lambda}$ from Table~\ref{Cv.Table} for each 
family separately we obtain the symmetry structures given in Table \ref{PottsXY.Table}.
According to this table, the symmetry pattern repeats itself every three $S_z$ sectors.
In particular, the full set of spatial IR's of Eqs.~(\ref{CubocRed.Eq}) and (\ref{IcosiRed.Eq})
is contained in any triad of subsequent $S_z$ sectors.
As can be seen from the above relations, this periodic pattern stems from the 120$^\circ$
3-sublattice symmetry structure of the coplanar states (the character
$\chi^\lambda(\phi) \sim \cos{(n \phi)}$ (with $\phi=0,2\pi/3,4\pi/3$) has a period of $n=3$).

In connection to a remark above, it is clear that we could have chosen here $\mathsf{L}=\mathsf{U}(1)$
instead of $\mathsf{C}_{\infty v}$. The corresponding stabilizer would then obviously be different from the above since 
none of the combinations of $\mathsf{S}_{bac}$, $\mathsf{S}_{acb}$ and $\mathsf{S}_{cba}$ with $\mathsf{U}(1)$ rotations
can bring $|c\rangle$ to itself. Nevertheless, the results from the two different choices of $\mathsf{L}$ are consistent
with each other.

We should finally emphasize a non-trivial feature which appears in both
Tables~\ref{PottsXY.Table} and \ref{PottsHeis.Table} and holds for each family separately.
Namely that the total number $m_\lambda$ of states (counting the degeneracy $d_\rho$ of spatial IR's)
for a given IR $\lambda$ of the spin group $\mathsf{L}$ is equal to the dimensionality $d_\lambda$
of this IR (i.e. $(2S + 1)$ for the Heisenberg case, see Table~\ref{Cv.Table} for the XY case)
times the ratio $|\mathsf{R}|/|\mathsf{H}_c|$. This feature can be proven by group theory alone
using $m_\lambda = \sum_\rho d_\rho~m_{\rho,\lambda}$, Eq.~(\ref{mgamma.Eq}) and the so-called
character orthogonality relation~\cite{GroupTheory}.

\end{document}